\documentclass[aps,twocolumn,showpacs]{revtex4}

\usepackage[dvips]{graphics}
\usepackage{fancyhdr}
\usepackage{setspace}
\usepackage{color}
\definecolor{gold}{rgb}{0.85,0.66,0}
\definecolor{dblue}{rgb}{0,0,0.8}

\begin{document}

\title{{\textcolor{gold}{Implementation of classical logic gates at
nano-scale level using magnetic quantum rings: A theoretical study}}}

\author{{\textcolor{dblue}{Santanu K. Maiti}}$^{\dag,\ddag,}$\footnote{{\bf 
Corresponding Author}: Santanu K. Maiti \\
$~~~$Electronic mail: santanu.maiti@saha.ac.in}}

\affiliation{$^{\dag}$Theoretical Condensed Matter Physics Division, 
Saha Institute of Nuclear Physics, 1/AF, Bidhannagar, Kolkata-700 064, 
India \\
$^{\ddag}$Department of Physics, Narasinha Dutt College, 129 Belilious 
Road, Howrah-711 101, India} 

\begin{abstract}
We explore the possibilities of designing classical logic gates at 
nano-scale level using magnetic quantum rings. A single ring is used 
for designing OR, NOT, XOR, XNOR and NAND gates, while AND and NOR gate 
responses are achieved using two such rings and in all the cases each 
ring is threaded by a magnetic flux $\phi$ which plays the central role 
in the logic gate operation. We adopt a simple tight-binding Hamiltonian
to describe the model where a magnetic quantum ring is attached to two
semi-infinite one-dimensional non-magnetic electrodes. Based on single 
particle Green's function formalism all the calculations which describe 
two-terminal conductance and current through the quantum ring are 
performed numerically. The analysis may be helpful in fabricating 
mesoscopic or nano-scale logic gates.
\end{abstract}


\maketitle

\section{Introduction}

Discovery of giant magnetoresistance effect in Fe/Cr magnetic 
multilayers~\cite{gmr} in 1988 ignited the idea of possibility to control 
and manipulate electron spin degree of freedom for storage and transfer of 
information as in conventional electronics. With the rapid progress in 
nanolithography and nanofabrication techniques~\cite{nanofab1,nanofab2} 
spin dependent transport at mesoscopic length scale is being paid much 
attention today from theoretical as well as experimental point of view 
due to its potential application in nanoscience and nanotechnology. 
Getting introduced in 1996 by S. Wolf `Spintronics'~\cite{wolf} has 
outgrown as one of the most enriched and sophisticated areas in condensed 
matter physics over the past two decades revolutionizing the concept of 
information storage technology. It holds future promises to integrate 
memory and logic into a single device. The key idea of designing spin 
dependent nano-electronic devices is based on the concept of quantum 
interference effect~\cite{imry1}, and it is generally preserved throughout
the sample having dimension smaller or comparable to the phase coherence
length. Therefore, ring type conductors or two path devices are ideal 
candidates where the effect of quantum interference can be 
exploited~\cite{bohm}. In such a ring shaped geometry, quantum 
interference effect can be controlled by several ways, and most
probably, the effect can be regulated significantly by tuning the
magnetic flux, the so-called Aharonov-Bohm (AB) flux, that threads the 
ring. A magnetic quantum ring penetrated by a magnetic flux $\phi$ 
yields a flux dependent spin transmission probability which may be 
useful for modeling of spin based logic gates and spin transistors.

In recent times, spin dependent transport through magnetic systems 
of ring shaped geometries has drawn much attention since these simple 
looking systems can be used to demonstrate several physical phenomena 
such as, many body correlation effect~\cite{manybody}, quantum phase 
transition~\cite{qphase}, resonant tunneling~\cite{resonant1,resonant2}, 
spin related conductance modulation~\cite{condmod}, spin 
filtering~\cite{filter}, spin detecting~\cite{detect}, etc. 
At the same time, much interest has also been shown in the study of spin 
based transport through a quantum ring in presence of an inhomogeneous
magnetic field. It provides us how such a system can used to make a device
of spin switch~\cite{frus,jia} and opens up the possibility of designing
spin filters~\cite{bird}, spin transistors~\cite{das}, and quantum 
information processing~\cite{imam}. In a very recent work Brataas 
{\em et al.}~\cite{brat} have shown that in
mesoscopic rings a spatially inhomogeneous spin-orbit (SO) interaction 
enhances the spin-interference effects and the transport mechanism can
be understood in terms of the AB physics with fictitious spin dependent 
magnetic fluxes. The inhomogeneous SO interaction controls and enhances 
spin injection in two-terminal rings significantly and these aspects can 
be used for quantum computation.

Following a brief introduction of spin dependent transport through a
magnetic quantum ring threaded by an AB flux $\phi$, in the present work
we will explore how such a quantum ring can be used for implementing 
classical logic gates. A single mesoscopic ring is used to design OR, NOT, 
XOR, XNOR and NAND gates, while AND and NOR gates are fabricated with the
help of two such quantum rings. For all these logic gates, AB flux $\phi$ 
enclosed by a ring plays the central role and it controls the interference 
condition of electronic waves passing through two arms of the ring. Within 
a non-interacting picture, a tight-binding framework is used to describe 
the model and all the calculations are done based on single particle 
Green's function technique~\cite{tho,san1,lee,san2,ando,san3,guo,san4}. 
There are also several other methods like mode matching 
techniques~\cite{modematch1,modematch2,modematch3}, transfer matrix 
method~\cite{transfer1,transfer2,transfer3,transfer4,transfer5}, etc., 
those are used to study spin dependent transport in low-dimensional 
model quantum systems. The logical operations are addressed by studying 
two-terminal conductance as a function of energy and current as a 
function of applied bias voltage. Our numerical analysis clearly supports 
the logical operations of the traditional macroscopic logic gates. To the 
best of our knowledge, the logic gate operations using such a simple 
magnetic quantum ring have not been described earlier in the literature.

The organization of the paper is as follows. With a brief introduction
given in Section I, in Section II we illustrate the theoretical 
formulation of spin dependent transport through a magnetic system 
sandwiched between two one-dimensional ($1$D) semi-infinite non-magnetic
(NM) metallic electrodes. The system between two electrodes can be anything 
like a $1$D magnetic chain, an array of quantum dots have finite magnetic
moments, a magnetic quantum ring, etc. In Section III, we present our 
numerical results which describe conductance-energy and current-voltage 
characteristics. At the end, summary of our results will be available in 
Section IV.

\section{Synopsis of the theoretical formulation} 

\begin{figure}[ht]
{\centering \resizebox*{6.5cm}{3.5cm}{\includegraphics{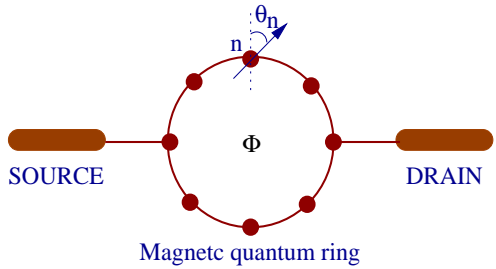}}\par}
\caption{(Color online). A magnetic quantum ring, penetrated by a
magnetic flux $\phi$, is attached symmetrically (upper and lower arms 
of the ring contain identical number of equally spaced atomic sites)
to two semi-infinite $1$D non-magnetic electrodes, viz, source and
drain. The filled red circles correspond to the positions of magnetic
sites in the ring.}
\label{ring}
\end{figure}

Let us start by referring to Fig.~\ref{ring}. A magnetic quantum ring
penetrated by an AB flux $\phi$ is attached symmetrically to two
semi-infinite $1$D non-magnetic metallic electrodes to form a bridge
system, the so-called electrode-conductor-electrode bridge. Filled 
red circles correspond to the positions of magnetic sites in the ring.
The strength of the localized magnetic moment associated with each 
magnetic site $n$ (say) is described by the parameter $h_n$ and its 
(magnetic moment) orientation is specified by the polar angle $\theta_n$ 
and azimuthal angle $\varphi_n$ in spherical polar coordinate system. 
Applying an external magnetic field, orientation of a local magnetic 
moment can be changed. The magnetic quantum ring with $N$ atomic sites 
is attached symmetrically to two $1$D semi-infinite non-magnetic metallic 
electrodes, namely, source and drain having chemical potentials $\mu_1$ 
and $\mu_2$ under the non-equilibrium condition when bias voltage is 
applied. Described by the discrete lattice model, the electrodes are 
assumed to be composed of infinite non-magnetic sites labeled as $0$, 
$-1$, $-2$, $\ldots$, $-\infty$ for the left electrode and $(N+1)$, 
$(N+2)$, $(N+3)$, $\ldots$, $\infty$ for the right one.

The Hamiltonian for the full system can be written as,
\begin{equation}
H=H_{R}+H_S+H_D+H_{SR}+H_{RD}
\label{equ1}
\end{equation}
where, $H_{R}$ represents the Hamiltonian for the magnetic quantum ring
(MQR). $H_S$ and $H_D$ correspond to the Hamiltonians for the source and
drain, respectively, and $H_{SR(RD)}$ is the Hamiltonian describing
the ring-electrode coupling strength.

The spin polarized Hamiltonian for the MQR can be written in 
non-interacting electron picture within the framework of tight-binding 
formulation in Wannier basis, using nearest-neighbor approximation as,
\begin{eqnarray}
H_{R} = \sum_{n=1}^N {\bf c_n^{\dagger} \left(\epsilon_0
-\vec{h_n}.\vec{\sigma} \right) c_n}
 + \sum_{i=1}^N 
{\bf \left(c_i^{\dagger}tc_{i+1} + h.c. \right)}
\label{equ2}
\end{eqnarray}
where, \\
${\bf c_n^{\dagger}}=\left(\begin{array}{cc}
c_{n \uparrow}^{\dagger} & c_{n \downarrow}^{\dagger} \end{array}\right);~~
{\bf c_n}=\left(\begin{array}{c}
c_{n \uparrow} \\
c_{n \downarrow}\end{array}\right)$ \\
${\bf \epsilon_0}=\left(\begin{array}{cc}
\epsilon_0 & 0 \\
0 & \epsilon_0 \end{array}\right);~~
{\bf t} e^{i\Theta} = t e^{i\Theta}\left(\begin{array}{cc}
1 & 0 \\
0 & 1 \end{array}\right)$ \\
${\bf \vec{h_n}.\vec{\sigma}} = h_n\left(\begin{array}{cc}
\cos \theta_n & \sin \theta_n e^{-i \varphi_n} \\
\sin \theta_n e^{i \varphi_n} & -\cos \theta_n \end{array}\right)$ \\
~\\
\noindent
First term of Eq.~\ref{equ2} represents the effective on-site energies of 
the atomic sites in the ring. $\epsilon_0$'s are the site energies, while 
the term ${\bf \vec{h_n}.\vec{\sigma}}$ describes the interaction of the 
spin ${\bf \sigma}$ of the injected electron to the localized on site 
magnetic moments. On site flipping of spins is described mathematically 
by this term. Second term describes the nearest-neighbor hopping strength 
between the sites of the quantum ring, modified due to the presence of AB 
flux $\phi$ which is incorporated by the term $\Theta=2\pi \phi/N \phi_0$.

Similarly, the Hamiltonian $H_{S(D)}$ for the two electrodes can be
written as,
\begin{equation}
H_{S(D)}=\sum_i {\bf c_i^{\dagger} \epsilon_{S(D)} c_i} + \sum_i
{\bf \left(c_i^{\dagger} t_{S(D)} c_{i+1} + h.c. \right)}
\label{equ3}
\end{equation}
where, $\epsilon_{S(D)}$'s are the site energies of source (drain) and
$t_{S(D)}$ is the hopping strength between the nearest-neighbor sites
of source (drain).

\noindent
Here also, \\
~\\
${\bf \epsilon_{S(D)}}=\left(\begin{array}{cc}
\epsilon_{S(D)} & 0 \\
0 & \epsilon_{S(D)} \end{array}\right);~~
{\bf t_{S(D)}}=\left(\begin{array}{cc}
t_{S(D)} & 0 \\
0 & t_{S(D)} \end{array}\right)$ \\
~\\
\noindent
The ring-electrode coupling Hamiltonian is described by,
\begin{equation}
H_{SR(RD)}= {\bf \left(c_{0(N)}^{\dagger} t_{SR(RD)} c_{1(N+1)} + 
h.c.\right)}
\label{equ4}
\end{equation}
where, $t_{SR(RD)}$ being the ring-electrode coupling strength.

In order to calculate the spin dependent transmission probabilities and 
current through the magnetic quantum ring, we use single particle Green's 
function technique. Within the regime of coherent transport and for 
non-interacting systems this formalism is well applied.

The single particle Green's function representing the full system for an 
electron with energy $E$ is defined as,
\begin{equation}
\bf{G}=(\bf{E}-\bf{H})^{-1}
\label{equ5}
\end{equation}
where,
\begin{equation}
{\bf{E}} = (\epsilon + i \eta) {\bf{I}}
\label{equ6}
\end{equation}
$\epsilon$ being the energy of the electron passing through the system.
$i \eta$ is a small imaginary term added to make the Green's function
$(\bf{G})$ non-hermitian.

Now $\bf{H}$ and $\bf{G}$ representing the Hamiltonian and the Green's 
function for the full system can be partitioned like~\cite{datta1,datta2},
\begin{equation}
\bf{H}=\left(\begin{array}{ccc}
\bf{H_{S}} & \bf{H_{SR}} & 0 \\
\bf{H_{SR}^\dag} & \bf{H_{R}} & \bf{H_{RD}}\\
0 & \bf{H_{RD}^\dag} & \bf{H_{D}}\\
\end{array} \right) 
\label{equ7}
\end{equation}
\begin{equation}
\bf{G}=\left(\begin{array}{ccc}
\bf{G_{S}} & \bf{G_{SR}} & 0 \\
\bf{G_{SR}^\dag} & \bf{G_{R}} & \bf{G_{RD}}\\
0 & \bf{G_{RD}^\dag} & \bf{G_{D}}\\
\end{array} \right) 
\label{equ8}
\end{equation}
where, $\bf{H_S}$, $\bf{H_R}$, and $\bf{H_D}$ represent the Hamiltonians 
(in matrix form) for source, quantum ring and  drain, respectively. 
$\bf{H_{SR}}$ and $\bf{H_{RD}}$ are the matrices for the Hamiltonians 
representing the ring-electrode coupling strength. Assuming that there 
is no coupling between the electrodes themselves, the corner elements of 
the matrices are zero. A similar definition goes for the Green's function 
matrix $G$ as well.

Our first goal is to determine $\bf{G_{R}}$ (Green's function for the 
ring only) which defines all physical quantities of interest. Following 
Eq.~\ref{equ5} and using the block matrix form of $\bf{H}$ and $\bf{G}$ 
the form of $\bf{G_{R}}$ can be expressed as,
\begin{equation}
\bf{G_{R}}=(\bf{E}-\bf{H_{R}}-\bf{\Sigma_{S}}-\bf{\Sigma_{D}})^{-1}
\label{equ9}
\end{equation}
where, $\bf{\Sigma_{S}}$ and  $\bf{\Sigma_{D}}$ represent the contact 
self-energies introduced to incorporate the effects of semi-infinite 
electrodes coupled to the system, and, they are expressed by the 
relations~\cite{datta1,datta2},
\begin{eqnarray}
\bf{\Sigma_{S}} & = & \bf{H_{SR}^{\dag} G_{S} H_{SR}} \nonumber \\
\bf{\Sigma_{D}} & = & \bf{H_{RD}^{\dag} G_{D} H_{RD}}
\end{eqnarray}
Thus the form of self-energies are independent of the nano-structure
itself through which transmission is studied and they completely
describe the influence of electrodes attached to the system. Now, the 
transmission probability $T_{\sigma \sigma^{\prime}}$ of an electron 
with energy $E$ is related to the Green's function as,
\begin{eqnarray}
T_{\sigma \sigma^{\prime}} & = & {\bf \Gamma}^{1}_{\bf{S}(\sigma \sigma)}
{\bf G}^{1N}_{\bf{r} (\sigma \sigma^{\prime})} 
{\bf G}^{N1}_{\bf{a} (\sigma^{\prime} \sigma)}
{\bf \Gamma}^{N}_{\bf{D}(\sigma^{\prime} \sigma^{\prime})} \nonumber \\
& = & {\bf \Gamma}^{1}_{\bf{S}(\sigma \sigma)}
|{\bf G}^{1N}_{ (\sigma \sigma^{\prime})}|^2
{\bf \Gamma}^{N}_{\bf{D}(\sigma^{\prime} \sigma^{\prime})}
\label{equ11}
\end{eqnarray}
where,
${\bf \Gamma}^{1}_{\bf{S}(\sigma \sigma)} = \langle 1 \sigma| 
{\bf \Gamma_S} | 1 \sigma \rangle $,
${\bf \Gamma}^{N}_{\bf{D}(\sigma^{\prime} \sigma^{\prime})} = 
\langle N \sigma^{\prime}| {\bf \Gamma_D} |N \sigma^{\prime} \rangle $
and ${\bf G}^{1 N}_{\sigma \sigma^{\prime}} =
\langle 1 \sigma| {\bf G} |N \sigma^{\prime} \rangle $.
Here, $\bf{G_{r}}$ and  $\bf{G_{a}}$ are the retarded and advanced single 
particle Green's functions (for the MQR only) for an electron with energy 
$E$. $\bf{\Gamma_{S}}$ and $\bf{{\Gamma_{D}}}$ are the coupling matrices, 
representing the coupling of the magnetic quantum ring to the source and 
drain, respectively, and they are defined by the relation~\cite{datta1,
datta2},
\begin{equation}
\bf{\Gamma_{S(D)}} = i[\Sigma^r_{S(D)} - \Sigma^{a}_{S(D)}]
\label{equ12}
\end{equation}
Here, $\bf{\Sigma^r_{S(D)}}$ and $\bf{\Sigma^a_{S(D)}}$ are the retarded 
and advanced self-energies, respectively, and they are conjugate to each 
other. It is shown by Datta {\em et al.} that the self-energy can be 
expressed as a linear combination of real and imaginary parts in the 
form,
\begin{equation}
{\bf{\Sigma^r_{S(D)}}} = {\bf\Lambda_{S(D)}} - i {\bf\Delta_{S(D)}}
\label{equ13}
\end{equation}
The real part of self-energy describes the shift of the energy levels
and the imaginary part corresponds to broadening of the levels. The
finite imaginary part appears due to incorporation of the semi-infinite
electrodes having continuous energy spectrum. Therefore, the coupling
matrices can be easily obtained from the self-energy expression and is
expressed as,
\begin{equation}
{\bf{\Gamma_{S(D)}}}=-2 {\bf Im} {\bf{(\Sigma_{S(D)})}}
\label{equ14}
\end{equation}
Considering linear transport regime, conductance $g_{\sigma 
\sigma^{\prime}}$ is obtained using Landauer formula~\cite{datta1,
datta2,land1,land2,land3},
\begin{equation}
g_{\sigma \sigma^{\prime}}=\frac{e^2}{h}T_{\sigma \sigma^{\prime}}
\label{equ15}
\end{equation}
Knowing the transmission probability $T_{\sigma \sigma^{\prime}}$ of
an electron injected with spin $\sigma$ and transmitted with spin
$\sigma^{\prime}$, the current $I_{\sigma \sigma^{\prime}}$ through the
system is obtained using Landauer-B\"{u}ttiker formalism. It is written
in the form~\cite{datta1,datta2,land1,land2,land3},
\begin{equation}
I_{\sigma \sigma^{\prime}} (V)= \frac{e}{h} \int \limits_{-\infty}^{+\infty} 
\left[f_S(E)-f_D(E)\right] T_{\sigma \sigma^{\prime}}(E)~dE
\label{equ16}
\end{equation}
where, $f_{S(D)}=f(E-\mu_{S(D)})$ gives the Fermi distribution function of 
the two electrodes having chemical potentials $\mu_{S(D)}=E_{F} \pm eV/2$. 
$E_F$ is the equilibrium Fermi energy and $V$ is the applied bias voltage. 
Here we make a realistic assumption that the entire voltage is dropped 
across the ring-electrode interfaces, and it is examined that under such 
an assumption the current-voltage ($I$-$V$) characteristics do not change 
their qualitative features significantly.

\section{Numerical results and discussion}

We start analyzing our results by mentioning the values of different 
parameters used for the numerical calculations. For a magnetic quantum 
ring, we set the on-site energy $\epsilon_0=0$ and nearest-neighbor 
hopping strength $t=3$. Magnitude of the local magnetic moment $h$, 
associated with each atomic site of the ring, is fixed at $0.5$. On
the other hand, for two non-magnetic electrodes the on-site energy
is taken as $\epsilon_S=\epsilon_D=0$ and nearest-neighbor hopping
integral is chosen as $t_S=t_D=4$. The equilibrium Fermi energy $E_F$ 
is fixed at $0$. For our illustrative purposes, we simplify the unit 
system by choosing $c=e=h=1$. Energy scale is fixed in unit of $t$.

In a bridge system (i.e., where a conductor is sandwiched between two
electrodes), transport properties are significantly influenced by the 
conductor-to-electrode coupling strength ($t_{SR(RD)}$). To emphasize 
it, we describe our results for the two limiting cases depending on
the coupling strength of the conductor (magnetic quantum ring) to the 
side attached NM electrodes. We define these two regimes as follows.
\vskip 0.1cm
\noindent
\underline{Case 1:} Weak-coupling limit
\vskip 0.1cm
\noindent
This regime is typically defined by the condition $t_{SR(RD)}<<t$. Here, 
we choose the values of the hopping parameters as $t_{SR}=t_{RD}=0.5$.

\vskip 0.1cm
\noindent
\underline{Case 2:} Strong-coupling limit
\vskip 0.1cm
\noindent
This limit is described by the condition $t_{SR(RD)} \sim t$. In this
regime we set the values of hopping strengths as $t_{SR}=t_{RD}=2.5$.

In order to understand the basic mechanisms of designing classical 
logic gates using magnetic quantum rings, let us first discuss the spin 
dependent transport through a magnetic quantum ring, penetrated by an 
AB flux $\phi$, which is symmetrically coupled to two non-magnetic 
metallic electrodes (for instance see Fig.~\ref{ring}). 

\subsection{A magnetic quantum ring}

\subsubsection{Conductance-energy characteristics}

\begin{figure}[ht]
{\centering \resizebox*{7.75cm}{10cm}{\includegraphics{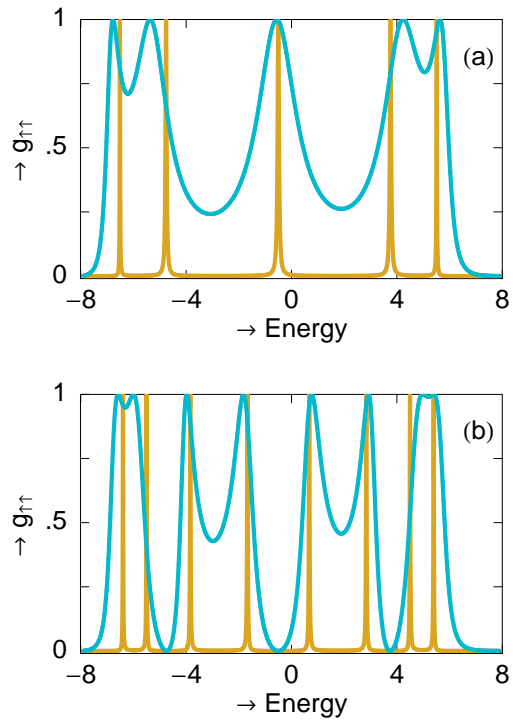}}\par}
\caption{(Color online). Up spin conductances ($g_{\uparrow \uparrow}$) as a 
function of energy $E$ for a magnetic quantum ring with $N=8$. The golden 
yellow and sky blue curves correspond to the weak- and strong-coupling 
cases, respectively. (a) $\phi=0$ and (b) $\phi=\phi_0/4$. Other parameters 
are as follows: $\varphi_n=0$ and $\theta_n=0$ for all sites $n$.
$g_{\uparrow \uparrow}$ is measured in unit of $e^2/h$.}
\label{ringcond}
\end{figure}

As representative examples, in Fig.~\ref{ringcond} we display the 
variations of up spin conductances $g_{\uparrow \uparrow}$ as a 
function of injecting electron energy $E$ for a magnetic quantum ring
considering $N=8$. The local moments associated with the magnetic 
atoms in two arms of the ring are aligned along $+Z$ direction i.e.,
$\theta_n=\varphi_n=0$ for all $n$. The golden yellow and sky blue 
lines represent
the results for the weak and strong ring-to-electrode coupling limits,
respectively. In the limit of weak-coupling when AB flux is not given to 
the ring, conductance shows fine resonant peaks (Fig.~\ref{ringcond}(a)) 
for some typical energy values, while for all other energies it almost 
drops to zero. At resonances, conductance $g_{\uparrow \uparrow}$
approaches to unity, and therefore, the transmission probability 
$T_{\uparrow \uparrow}$ becomes $1$ since we get the relation 
$g_{\uparrow \uparrow}=T_{\uparrow \uparrow}$ from the Landauer 
conductance formula (see Eq.~\ref{equ15}) in our chosen unit $e=h=1$.
$T_{\uparrow \uparrow}=1$ reveals a ballistic transmission through the
MQR. Each resonant peak in the conductance spectrum is associated with
a particular energy eigenvalue of the ring. Thus, from the
conductance-energy spectrum, nature of the energy eigenvalues of the 
quantum ring can be directly implemented. The sharpness of resonant 
peaks drastically changes in the limit of strong ring-electrode coupling 
which is clearly visible from the sky blue curve of Fig.~\ref{ringcond}(a). 
In this strong-coupling limit, all resonant peaks get broadened where 
the contribution for the broadening comes from the imaginary parts 
of the self-energies $\Sigma_S$ and $\Sigma_D$~\cite{datta1,datta2}.
These results predict that in the limit of weak-coupling a fine tuning 
in energy scale is required to get spin transmission across the ring,
while for the strong-coupling limit since conduction takes place in
a wide range of energy, fine tuning is not necessary to get electron
conduction. This coupling effect provides an important signature in
the study of electron transport and the effect becomes much clearer 
from our current-voltage ($I$-$V$) characteristics. In the presence of
AB flux $\phi$, more resonant peaks appear in the conductance spectrum
compared to the case when $\phi=0$. The results are shown in 
Fig.~\ref{ringcond}(b) where we set $\phi=\phi_0/4$. Appearance of
more resonant peaks is associated with the existence of more discrete 
energy levels which is caused by removal of energy degeneracies in 
the presence of $\phi$.

All the results presented above (Fig.~\ref{ringcond}) are associated 
with the variation of only up spin conductance $g_{\uparrow \uparrow}$ 
as a function of energy $E$. The conductance-energy spectrum for a down 
spin electron is exactly mirror symmetric to the spectrum observed for 
an up spin electron, and accordingly, we do not plot the results further 
for down spin electrons. This mirror symmetric like feature is observed 
only when we set the site energy $\epsilon_0$ of the MQR to zero. In 
addition, it is also important to note that for this MQR where all moments 
are aligned along $+Z$ direction, no spin flip transmission takes place 
i.e., $T_{\uparrow \downarrow}=T_{\downarrow \uparrow}=0$. The explanation 
of zero transmission probability for spin flipping is given as follow. Spin 
flip occurs due to the presence of the term $\vec{h}.\vec{\sigma}$ in the 
Hamiltonian (see Eq.~\ref{equ2}), $\vec{\sigma}$ being the Pauli spin 
matrix with components $\sigma_x$, $\sigma_y$ and $\sigma_z$ for the 
injecting electron. The spin flipping is caused because of the operators 
$\sigma_+ (=\sigma_x + i\sigma_y)$ and $\sigma_- (=\sigma_x - i\sigma_y)$, 
respectively. For the local magnetic moments oriented along $\pm$ $Z$ axes, 
$\vec{h}.\vec{\sigma}$ $(= h_x\sigma_x + h_y\sigma_y + h_z\sigma_z)$ becomes 
equal to $h_z\sigma_z$. Accordingly, the Hamiltonian does not contain 
$\sigma_x$ and $\sigma_y$ and so as $\sigma_+$ and $\sigma_-$, which 
provides zero flipping for up or down orientation of magnetic moments. 

\subsubsection{Variation of conductance with AB flux $\phi$}

\begin{figure}[ht]
{\centering \resizebox*{7.55cm}{4.75cm}
{\includegraphics{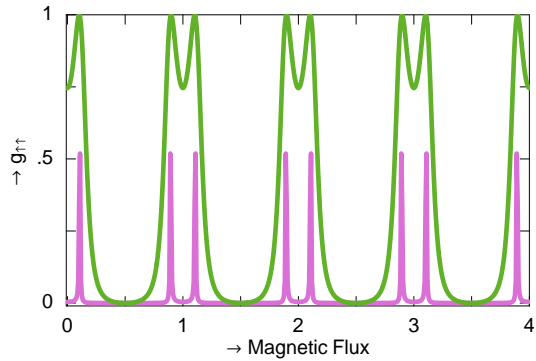}}\par}
\caption{(Color online). Up spin conductances $g_{\uparrow \uparrow}$, 
calculated at the typical energy $E=0$, as a function of magnetic flux 
$\phi$ for a magnetic quantum ring with $N=8$. The magenta and green curves 
correspond to the weak- and strong-coupling cases, respectively. Other 
parameters are as follows: $\varphi_n=0$ and $\theta_n=0$ for all magnetic 
sites $n$.}
\label{ringcondphi}
\end{figure}

In Fig.~\ref{ringcondphi} we show the dependence of up spin conductances
as a function of AB flux $\phi$ for a magnetic quantum ring considering
$N=8$ where all moments are aligned along $+Z$ direction. The magenta
and green curves correspond to the up spin conductances for the weak-
and strong-coupling limits, respectively, and these conductances are
determined at the typical energy $E=0$. It is observed that the up
spin conductance in the case of weak-coupling is significantly reduced
compared to the strong-coupling one, and, both for these two coupling
cases up spin conductance varies periodically with $\phi$ showing 
$\phi_0$ ($=1$ in our chosen unit $c=e=h=1$) flux-quantum periodicity.
Quite interestingly we notice that at $\phi=n \phi_0/2$, where $n$ is
an odd integer, conductance drops exactly to zero. This vanishing 
behavior can be explained as follows.

For a symmetrically connected ring, the wave functions passing through 
the upper and lower arms of the ring are given by,
\begin{eqnarray}
\psi_1 & = & \psi_0 e^{\frac{ie}{\hbar c} \int \limits_{\gamma_1} 
\vec{A}.\vec{dr}} \nonumber \\
\psi_2 & = & \psi_0 e^{\frac{ie}{\hbar c} \int \limits_{\gamma_2} 
\vec{A}.\vec{dr}} 
\label{equ17}
\end{eqnarray}
where, $\gamma_1$ and $\gamma_2$ are used to indicate two different
paths of electron propagation along the two arms of the ring. $\psi_0$ 
denotes the wave function in the absence of magnetic flux $\phi$ and it is 
same for both upper and lower arms as the ring is symmetrically coupled 
to the electrodes. $\vec{A}$ is the vector potential associated with the 
magnetic field $\vec{B}$ by the relation $\vec{B}= \vec{\nabla} \times 
\vec{A}$. Hence the probability amplitude of finding the electron passing 
through the ring can be calculated as,
\begin{equation}
|\psi_1 + \psi_2|^2 = 2|\psi_0|^2 + 2|\psi_0|^2 \cos \left({\frac{2\pi 
\phi}{\phi_0}}\right)
\label{equ18}
\end{equation}
where, $\phi = \oint \vec{A}.\vec{dr} = \int \int \vec{B}.\vec{ds}$
is the flux enclosed by the ring.

It is clearly observed from Eq.~\ref{equ18} that at $\phi=n\phi_0/2$,
where $n$ stands for an odd integer, the transmission probability of an 
electron exactly drops to zero irrespective of the energy $E$. For any
other values of $\phi$, electron conduction no longer vanishes for the 
entire energy range and we get resonant peaks at the appropriate energy 
values. On the other hand, for an asymmetrically connected MQR, the 
vanishing behavior of conductance for the entire energy range at odd 
integer multiples of $\phi_0/2$ will not be observed, since in this 
case the wave amplitudes for the waves passing through the upper and
lower arms are not identical to each other. These aspects can be utilized 
to design all possible logic gates which we will describe in the 
forthcoming sub-sections. 

\subsubsection{Current-voltage characteristics}

All the basic features of spin dependent transport obtained from 
conductance-energy spectra of the magnetic quantum ring can be understood 
in a better way from our current-voltage characteristics. The current
passing through the ring is determined by integrating the transmission
function $T_{\sigma \sigma^{\prime}}$ according to Eq.~\ref{equ16}, where 
the transmission function varies exactly similar to that of the conductance 
spectrum apart from a scale factor $e^2/h$ (see Eq.~\ref{equ15}), which is 
equal to $1$ in our chosen unit system $e=h=1$.

\begin{figure}[ht]
{\centering \resizebox*{7.75cm}{9cm}{\includegraphics{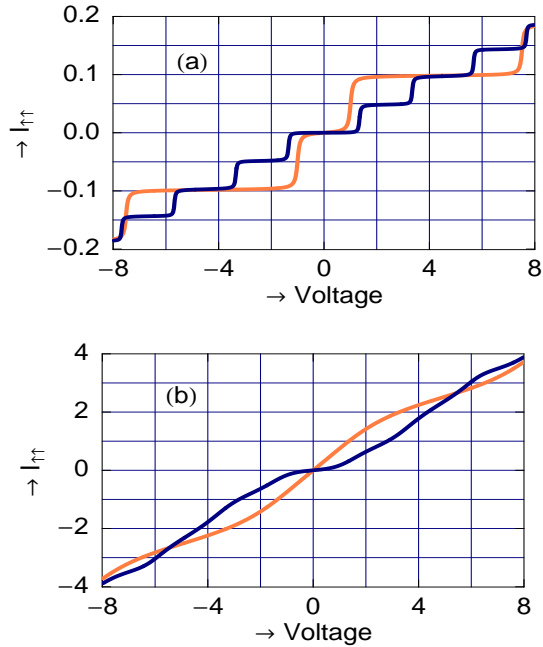}}\par}
\caption{(Color online). Up spin currents ($I_{\uparrow \uparrow}$) as 
a function of bias voltage $V$ for a magnetic quantum ring with $N=8$. 
The reddish-yellow and blue lines correspond to $\phi=0$ and $\phi_0/4$, 
respectively. (a) Weak-coupling limit and (b) strong-coupling limit. 
Other parameters are as follows: $\varphi_n=0$ and $\theta_n=0$ for 
all magnetic sites $n$. $I$ and $V$ are measured in units of $te/h$ and
$t/e$, respectively.}
\label{ringcurr}
\end{figure}

As illustrative examples, in Fig.~\ref{ringcurr} we present the variations
of up spin currents $I_{\uparrow \uparrow}$ as a function of applied bias
voltage $V$ for a symmetrically connected magnetic quantum ring where all
moments are aligned along $+Z$ direction. Here we fix $N=8$ so that each 
arm of the ring contains $3$ magnetic sites. The reddish-yellow and blue 
curves represent the currents for $\phi=0$ and $\phi_0/4$, respectively.
In the limit of weak-coupling, current shows step-like behavior as a 
function of bias voltage $V$ (Fig.~\ref{ringcurr}(a)), where each step is 
associated with a sharp resonant peak in the conductance-energy spectrum. 
With the increment of bias voltage $V$, electrochemical potentials $\mu_1$ 
and $\mu_2$ in the two electrodes shift gradually and eventually cross one 
of the quantized energy levels of the MQR, and therefore, a current step 
appears in the current-voltage characteristic curve. The effect of AB flux 
$\phi$ is quite interesting and it can be understood 
clearly by noting the curves presented in Fig.~\ref{ringcurr}(a). In
presence of $\phi$ ($\phi=\phi_0/4$) more steps appear in the 
current-voltage characteristic curve (blue line) associated with the
conductance-energy spectrum, compared to the case when $\phi$ is set
to zero (reddish-yellow line). With this feature it also important to
note that the typical bias voltage where electron starts to conduct
through the electrode-ring-electrode bridge, the so-called threshold
voltage $V_{th}$, can be regulated significantly through the AB flux
$\phi$. As $\phi$ increases towards $\phi_0/2$ from $0$ value, $V_{th}$
gets increased gradually, and, at the typical value of $\phi=\phi_0/2$,
$V_{th}$ is maximum where electron conduction is stopped for the complete
range of bias voltage. This phenomenon can be utilized in designing 
nano-electronic devices. 

The step-like behavior almost disappears in the case of strong
ring-to-electrode coupling limit, as shown in Fig.~\ref{ringcurr}(b),
where the reddish-yellow and blue curves correspond to the identical
meaning as in Fig.~\ref{ringcurr}(a). Here the current amplitude is 
very high compared to the weak-coupling limit. This continuous like
feature with large current amplitude can be easily explained from the
conductance-energy spectrum discussed earlier. Thus, in short we can
say that for a fixed bias voltage, current amplitude can be controlled
by tuning the ring-electrode coupling strength.

In the following sub-sections we will try to explore how these features 
of spin dependent transmission through a MQR can be implemented to 
fabricate several classical logic gates. We will describe two-input
logic gates where the inputs are associated with externally applied 
magnetic fields through which we can tune the orientation of local
magnetic moments in the magnetic atomic sites. Throughout our
discussion we mention the two inputs in terms of polar angles 
(expressed by the symbol $\theta$) of two magnetic moments associated 
with applied gate magnetic fields. For all logic gate operations AB 
flux $\phi$ plays the central role and we fix it at $\phi_0/2$ i.e.,
$0.5$ in our chosen unit system.

\subsection{OR gate}

Let us start with OR gate response. The schematic view of a magnetic 
\begin{figure}[ht]
{\centering \resizebox*{6.5cm}{3cm}{\includegraphics{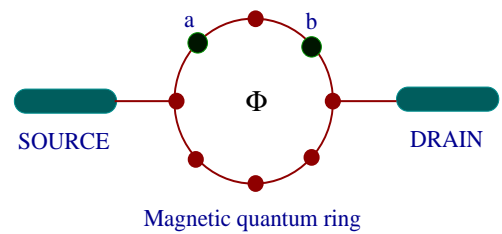}}\par}
\caption{(Color online). A magnetic quantum ring, penetrated by an AB
flux $\phi$, is attached symmetrically to two semi-infinite $1$D 
non-magnetic metallic electrodes. The magnetic sites $a$ and $b$
are subject to external gate magnetic fields through which the
orientations of local magnetic moments in these two sites are tuned.}
\label{or}
\end{figure}
quantum ring that can be used as an OR gate is shown in Fig.~\ref{or}.
The ring, penetrated by an AB flux $\phi$, is symmetrically coupled to 
two semi-infinite $1$D non-magnetic metallic electrodes, namely, source
and drain. Two magnetic sites $a$ and $b$ in upper arm of the ring are 
subject to two external gate magnetic fields through which the
orientations $\theta_a$ and $\theta_b$ of local magnetic moments 
associated with the respective sites $a$ and $b$ are controlled, and 
these two ($\theta_a$ and $\theta_b$) are treated as two inputs of 
the OR gate. Quite interestingly we observe that, {\em at $\phi=\phi_0/2$ 
a high output current ($1$) (in the logical sense) appears if one or 
both the inputs to the gate are high ($1$), while if neither input is 
high ($1$), a low output current ($0$) appears.} This phenomenon is the
so-called OR gate response and here we address it by studying 
conductance-energy and current-voltage characteristics as functions of 
magnetic flux and gate magnetic fields.

\begin{figure}[ht]
{\centering \resizebox*{7.75cm}{8cm}{\includegraphics{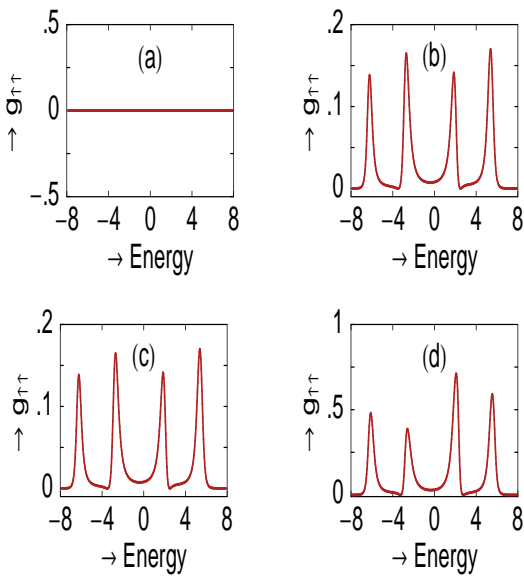}}\par}
\caption{(Color online). OR gate response. $g_{\uparrow \uparrow}$-$E$ 
curves for a magnetic quantum ring with $N=8$ in the limit of 
strong-coupling. (a) $\theta_a=\theta_b=0$, (b) $\theta_a=0$ and
$\theta_b=\pi$, (c) $\theta_a=\pi$ and $\theta_b=0$ and (d) 
$\theta_a=\theta_b=\pi$. Other parameters are as follows: $\phi=\phi_0/2$, 
$\theta_n=0$ for all magnetic sites $n$ except $n=a$ and $b$, and 
$\varphi_n=0$ for all sites $n$.}
\label{orcond}
\end{figure}

As illustrative examples in Fig.~\ref{orcond} we plot up spin conductances
$g_{\uparrow \uparrow}$ as a function of injecting electron energy $E$
for a magnetic quantum ring with $N=8$ in the limit of strong-coupling,
where (a), (b), (c) and (d) correspond to four different cases of two 
input signals. When both the two inputs are low ($0$) i.e., 
$\theta_a=\theta_b=0$, up spin conductance vanishes for the entire energy 
range (Fig.~\ref{orcond}(a)) which reveals no electron conduction through 
the ring. This vanishing behavior is clearly explained from our previous 
discussion where we have shown that, the transmission probability of an 
electron across a symmetrically connected ring (upper and lower lower 
arms are identical in nature) drops exactly to zero at the typical flux
$\phi=\phi_0/2$. With this argument we can justify the vanishing nature
of up spin conductance for the particular case when $\theta_a=\theta_b=0$.
Since in such a case the upper and lower arms are exactly identical to 
each other as we set $\theta_n=0$ for all $n$ except $a$ and $b$, and 
accordingly, zero transmission probability is achieved for the entire 
range of energy $E$. On the other hand, when the symmetry between the 
two arms of the ring is broken by applying an external magnetic field 
to anyone of the two input gates i.e., $\theta_a=0$ and $\theta_b=\pi$ 
(Fig.~\ref{orcond}(b)) or $\theta_a=\pi$ and $\theta_b=0$ 
(Fig.~\ref{orcond}(c)) or by applying external magnetic fields in both 
the two input gates i.e., $\theta_a=\theta_b=\pi$ (Fig.~\ref{orcond}(d)), 
up spin conductance shows resonant peaks for some particular energies 
associated with the energy eigenvalues of the magnetic quantum ring. 
\begin{figure}[ht]
{\centering \resizebox*{7.75cm}{8cm}{\includegraphics{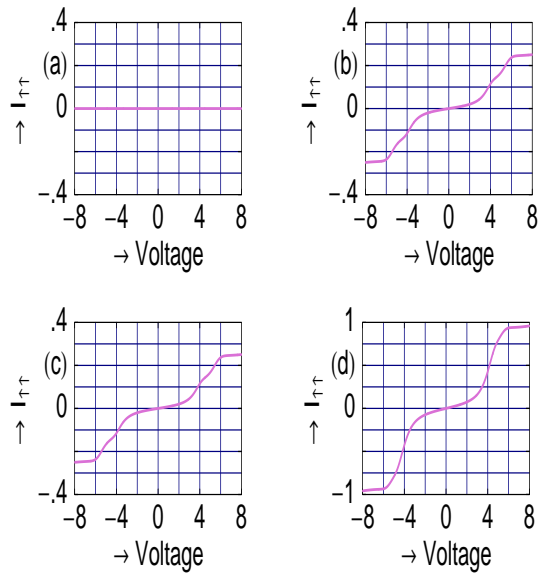}}\par}
\caption{(Color online). OR gate response. $I_{\uparrow \uparrow}$-$V$ 
curves for a magnetic quantum ring with $N=8$ in the limit of 
strong-coupling. (a) $\theta_a=\theta_b=0$, (b) $\theta_a=0$ and
$\theta_b=\pi$, (c) $\theta_a=\pi$ and $\theta_b=0$ and (d) 
$\theta_a=\theta_b=\pi$. Other parameters are as follows: $\phi=\phi_0/2$, 
$\theta_n=0$ for all magnetic sites $n$ except $n=a$ and $b$, and 
$\varphi_n=0$ for all sites $n$.}
\label{orcurr}
\end{figure}
In addition we also observe that in the cases where anyone of the two 
inputs to the gate is high and other is low, the height of the resonant 
peaks gets reduced compared to the case where both inputs are high. This
is solely due to the effect of quantum interference among the electronic
waves passing through two arms of the ring. From these conductance-energy 
spectra we can predict that electron conduction through the ring takes 
place when anyone or both the inputs to the gate are high ($1$), while for
the case where both inputs are low electron conduction is no longer 
possible. It emphasizes the OR gate behavior. In the above conductance-energy
spectra (Figs.~\ref{orcond}(a)-(d)), we display the variation of up spin 
conductance as a function of energy depending on the four different choices 
of two input signals. An exactly similar kind of behavior (OR gate response) 
will be also observed for down spin electrons, where conductance-energy 
spectrum gets mirror symmetric like nature compared to up spin electrons.
But, no spin flip transmission in these four different configurations 
takes place i.e., $T_{\uparrow \downarrow}$ and $T_{\downarrow \uparrow}$
are always zero as the moments are aligned along $\pm Z$ directions.

Following the above conductance-energy spectra now we describe the 
current-voltage characteristics. In Fig.~\ref{orcurr} we plot the 
variations of up spin currents $I_{\uparrow \uparrow}$ as a function of
applied bias voltage $V$ for the magnetic quantum ring ($N=8$) in the 
limit of strong-coupling, where (a), (b), (c) and (d) represent the results
for the four different choices of two inputs $\theta_a$ and $\theta_b$. 
When both inputs are low i.e., $\theta_a=\theta_b=0$, current vanishes
for the entire range of bias voltage $V$ (Fig.~\ref{orcurr}(a)). This is 
clearly explained from Fig.~\ref{orcond}(a), since current is evaluated by 
integrating the transmission function. While, in other three cases of 
two input signals i.e., $\theta_a=0$ and $\theta_b=\pi$ 
(Fig.~\ref{orcurr}(b)), $\theta_a=\pi$ and $\theta_b=0$ 
(Fig.~\ref{orcurr}(c)) and $\theta_a=\theta_b=\pi$ (Fig.~\ref{orcurr}(d)), 
up spin current shows a quite continuous variation with the bias voltage 
$V$, following the conductance-energy spectra (Figs.~\ref{orcond}(b)-(d)).
\begin{table}[ht]
\begin{center}
\caption{OR gate behavior in the limit of strong-coupling. The typical
current amplitude is determined at the bias voltage $V=6.26$.}
\label{ortable}
~\\
\begin{tabular}{|c|c|c|}
\hline \hline
Input-I ($\theta_a$) & Input-II ($\theta_b$) & Current \em{(I)} \\ \hline
$0$ & $0$ & $0$ \\ \hline 
$0$ & $\pi$ & $0.243$ \\ \hline 
$\pi$ & $0$ & $0.243$ \\ \hline
$\pi$ & $\pi$ & $0.940$ \\ \hline
\end{tabular}
\end{center}
\end{table}
Associated with the quantum interference effect among the two arms of the
ring, the larger current amplitude for a fixed bias voltage in the typical 
case where two inputs are high (Fig.~\ref{orcurr}(d)) compared to the
cases where one input is high and other is low (Figs.~\ref{orcurr}(b)-(c)) 
is clearly understood. These characteristics demonstrate that a magnetic 
quantum ring can be used as an OR gate. 

To be more precise, we make a quantitative estimate of the typical current 
amplitude, given in Table~\ref{ortable}, where the current amplitude is
measured at the bias voltage $V=6.26$. It shows that when both inputs are 
high ($\pi$), current gets the value $0.94$ and for the cases where anyone 
of the two inputs is high and other is low ($0$), current becomes $0.243$. 
On the other hand, current is zero for the particular case where both 
inputs are low. These aspects reveal the OR gate response in a magnetic
quantum ring.

\subsection{AND gate}

\begin{figure}[ht]
{\centering \resizebox*{7.7cm}{2.7cm}{\includegraphics{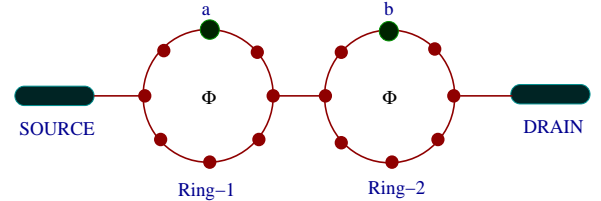}}\par}
\caption{(Color online). Two directly coupled magnetic quantum rings, 
where each ring is penetrated by an AB flux $\phi$, is attached 
symmetrically to two semi-infinite $1$D non-magnetic metallic 
electrodes. The magnetic sites $a$ and $b$ are subject to external 
magnetic fields through which the orientations of local magnetic 
moments in these two sites are tuned.}
\label{and}
\end{figure}

To design an AND logic gate we use two similar magnetic quantum rings
those are directly coupled to each other via a single bond. The schematic
view of the double quantum ring that can be used as an AND gate is presented
in Fig.~\ref{and}, where individual rings are penetrated by an AB flux 
$\phi$. The double quantum ring is then attached symmetrically to two 
semi-infinite $1$D metallic electrodes, namely, source and drain. Two 
magnetic sites $a$ and $b$ in upper arms of the two rings are subject 
to two external gate magnetic fields through which the orientations of 
local magnetic moments $\theta_a$ and $\theta_b$ in sites $a$ and $b$ 
are controlled. We consider them ($\theta_a$ and $\theta_b$) as the two 
\begin{figure}[ht]
{\centering \resizebox*{7.75cm}{8cm}{\includegraphics{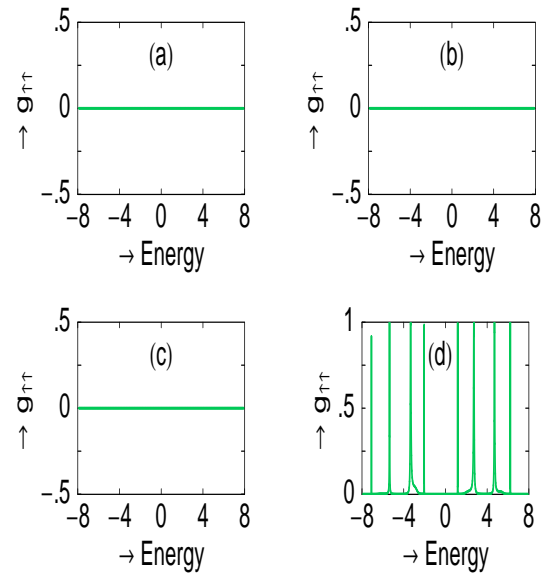}}\par}
\caption{(Color online). AND gate response. $g_{\uparrow \uparrow}$-$E$ 
curves in the strong-coupling limit for a double quantum ring where
each ring contains $8$ magnetic sites. (a) $\theta_a=\theta_b=0$, 
(b) $\theta_a=0$ and $\theta_b=\pi$, (c) $\theta_a=\pi$ and $\theta_b=0$ 
and (d) $\theta_a=\theta_b=\pi$. Other parameters are as follows: 
$\phi=\phi_0/2$, $\theta_n=0$ for all magnetic sites $n$ except $n=a$ and 
$b$, and $\varphi_n=0$ for all sites $n$.}
\label{andcond}
\end{figure}
inputs of the AND gate. We will show that, {\em at the typical flux 
$\phi=\phi_0/2$, a high output current ($1$) (in the logical sense) 
appears only if both the two inputs to the gate are high ($1$), while 
if neither or only one input to the gate is high ($1$), a low output 
current ($0$) results.} It is the so-called AND gate response and we 
investigate it by studying conductance-energy and current-voltage
characteristics.

To explore AND gate response, first we describe conductance-energy 
characteristics. In Fig.~\ref{andcond} we show the variations of up 
spin conductances ($g_{\uparrow \uparrow}$) as a function of energy 
$E$ in the limit of strong-coupling for a double quantum ring with $16$ 
($=2\times 8$) magnetic sites, where (a), (b), (c) and (d) represent
the results for different choices of input signals $\theta_a$ and
$\theta_b$.
\begin{figure}[ht]
{\centering \resizebox*{7.75cm}{8cm}{\includegraphics{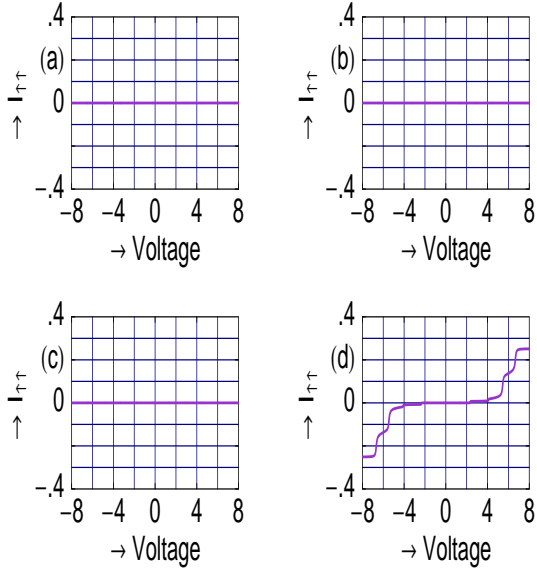}}\par}
\caption{(Color online). AND gate response. $I_{\uparrow \uparrow}$-$V$ 
curves in the limit of strong-coupling for a double quantum ring where
each ring contains $8$ magnetic sites. (a) $\theta_a=\theta_b=0$, (b) 
$\theta_a=0$ and $\theta_b=\pi$, (c) $\theta_a=\pi$ and $\theta_b=0$ and (d) 
$\theta_a=\theta_b=\pi$. Other parameters are as follows: $\phi=\phi_0/2$, 
$\theta_n=0$ for all magnetic sites $n$ except $n=a$ and $b$, and 
$\varphi_n=0$ for all sites $n$.}
\label{andcurr}
\end{figure}
It is observed that when both $\theta_a$ and $\theta_b$ are identical to
zero (low), up spin conductance vanishes throughout the energy range
(Fig.~\ref{andcond}(a)). The reason is that, by making two inputs to zero
the upper and lower arms of individual rings are similar in nature, and 
therefore, their contributions to the transmission probability disappear
at the typical flux $\phi=\phi_0/2$. A similar kind of vanishing behavior 
of up spin conductance in the complete energy band is also observed for 
the two other cases where anyone of the two inputs is high and other is 
low. The results are shown in Figs.~\ref{andcond}(b) and (c), where we
set $\theta_a=0$, $\theta_b=\pi$ and $\theta_a=\pi$, $\theta_b=0$,
respectively. In these two choices of input signals, symmetry between 
the upper and lower arms of two rings is not broken simultaneously.
When the symmetry is broken in one ring by applying a gate magnetic field, 
it (symmetry) is preserved in the other ring and vice versa. Now, for 
the asymmetric ring we get non-vanishing transmission probability,
while for the symmetric ring transmission amplitude becomes zero, and
therefore, as a combined effect we get vanishing transmission amplitude
since two rings are directly coupled to each other. The non-zero value of
up spin conductance is obtained only when the symmetry of the two rings 
are broken individually by applying external gate magnetic fields in the 
upper arms of two rings i.e., $\theta_a=\theta_b=\pi$ 
(Fig.~\ref{andcond}(d)). Thus, from the above conductance-energy spectra
we can predict that the electron conduction through the double quantum 
ring is possible only when both the inputs to the gate are high ($\pi$), 
while if neither or anyone input to the gate is high, no electron 
conduction takes place. These features are associated with traditional
AND gate response. 

Now we go for current-voltage characteristics to reveal AND gate response
in a double quantum ring. As representative examples, in Fig.~\ref{andcurr} 
we plot up spin currents $I_{\uparrow \uparrow}$ as a function of applied
bias voltage $V$ for a double quantum ring in the strong-coupling limit,
where (a), (b), (c) and (d) correspond to four different cases of two
input signals. When we put two inputs identically to low value i.e., 
$\theta_a=\theta_b=0$, up spin current becomes zero (Fig.~\ref{andcurr}(a)) 
for the full width of applied bias voltage $V$. An exactly similar vanishing 
behavior of up spin current is also available for the other two cases 
of input signals, where only one input is high and other is low. The
results are presented in Figs.~\ref{andcurr}(b) and (c). 
\begin{table}[ht]
\begin{center}
\caption{AND gate behavior in the limit of strong-coupling. The typical
current amplitude is determined at the bias voltage $V=6.26$.}
\label{andtable}
~\\
\begin{tabular}{|c|c|c|}
\hline \hline
Input-I ($\theta_a$) & Input-II ($\theta_b$) & Current \em{(I)} \\ \hline
$0$ & $0$ & $0$ \\ \hline 
$0$ & $\pi$ & $0$ \\ \hline 
$\pi$ & $0$ & $0$ \\ \hline
$\pi$ & $\pi$ & $0.145$ \\ \hline
\end{tabular}
\end{center}
\end{table}
For these three cases of the two input signals, the vanishing behavior
of up spin current can be easily understood from the conductance-energy 
spectra given in Figs.~\ref{andcond}(a)-(c), since current is determined
by integrating the transmission function. The finite non-zero value of
up spin current is available only for the typical case where both the 
two inputs are high i.e., $\theta_a=\theta_b=\pi$ (Fig.~\ref{andcurr}(d)),
following the conductance curve (Fig.~\ref{andcond}(d)). From this 
current-voltage curve we see that the up spin current shows non-zero
value beyond a finite value of bias voltage $V$, the so-called threshold
voltage $V_{th}$. This threshold voltage can be regulated by tuning the
ring-electrode coupling strength as well as by controlling the size of
the magnetic quantum ring. This characteristic provides an important 
signature in designing nano-electronic devices. These results support 
AND gate response.

In the same fashion as earlier here we also make a quantitative estimate
for the typical current amplitude as given in Table~\ref{andtable}, where
the typical current amplitude is measured at the bias voltage $V=6.26$.
It shows that the current gets the value $0.145$ only when both the two
inputs are high, while for all other cases i.e, where neither input is
high or only one is high, current drops exactly zero. It simplifies the
AND gate behavior in a double quantum ring.

\subsection{NOT gate}

\begin{figure}[ht]
{\centering \resizebox*{6.5cm}{3.5cm}{\includegraphics{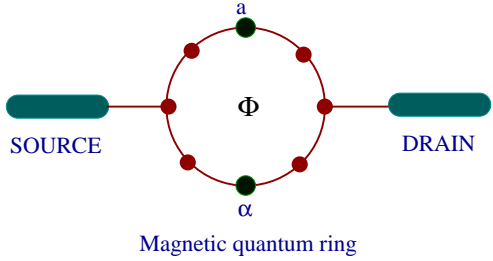}}\par}
\caption{(Color online). A magnetic quantum ring, threaded by an AB flux
$\phi$, is attached symmetrically to two semi-infinite $1$D non-magnetic
metallic electrodes. The magnetic sites $a$ and $\alpha$ are subject to 
two external gate magnetic fields through which the orientations of local
magnetic moments in these two sites are tuned.}
\label{not}
\end{figure}

Next we discuss NOT gate operation in a magnetic quantum ring. Schematic 
view for the operation of a NOT gate by using a single ring is shown in 
Fig.~\ref{not}, where the ring is attached symmetrically to two 
semi-infinite $1$D metallic electrodes, viz, source and drain, and
it is subject to an AB flux $\phi$. Applying gate magnetic fields in the 
magnetic sites named as $a$ and $\alpha$ in upper and lower arms of the 
ring we tune the directions of local magnetic moments $\theta_a$ and 
$\theta_{\alpha}$ in these two respective sites. Keeping $\theta_{\alpha}$ 
to a fixed value, we change $\theta_a$ properly to achieve the NOT gate 
operation. This $\theta_a$ is regarded as the input of our NOT gate. We 
will verify that, {\em at the typical flux $\phi=\phi_0/2$, a high output 
current ($1$) (in the logical sense) appears if the input to the gate is 
low ($0$), while a low output current ($0$) appears when the input to the 
gate is high ($1$).} This phenomenon is the so-called NOT gate behavior,
and we will explore it following the same prescription as earlier.

\begin{figure}[ht]
{\centering \resizebox*{7.75cm}{9cm}{\includegraphics{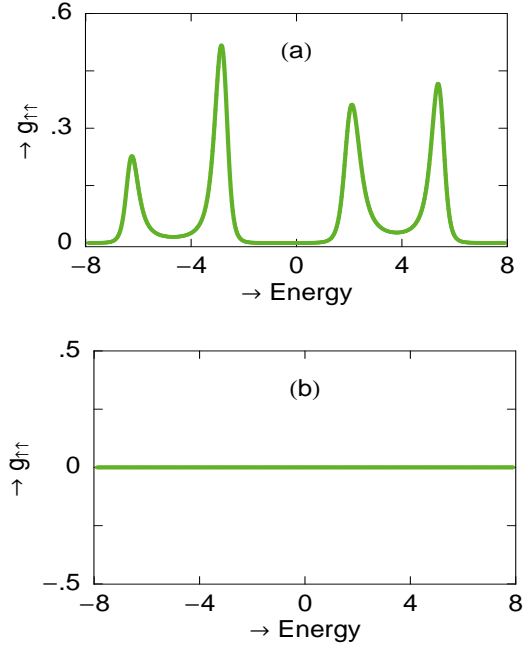}}\par}
\caption{(Color online). NOT gate response. $g_{\uparrow \uparrow}$-$E$ 
curves for a magnetic quantum ring with $N=8$ and $\theta_{\alpha}=\pi$
in the limit of strong-coupling. (a) $\theta_a=0$ and (b) $\theta_a=\pi$. 
Other parameters are as follows: $\phi=\phi_0/2$, $\theta_n=0$ for all 
magnetic sites $n$ except $n=a$ and $\alpha$, and $\varphi_n=0$ for all 
sites $n$.}
\label{notcond}
\end{figure}

To describe NOT gate operation let us start with the conductance-energy 
characteristics. In Fig.~\ref{notcond} we show the variations of up spin 
conductances ($g_{\uparrow \uparrow}$) as a function of injecting electron 
energy ($E$) for a typical magnetic quantum ring considering $N=8$ in 
the limit of strong-coupling, where (a) and (b) illustrate the results 
for two different choices of the input signal $\theta_a$. Throughout this 
logical operation we fix $\theta_{\alpha}$ to $\pi$ i.e., the moment in 
site $\alpha$ is oriented along the $-Z$ direction. From our results we
see that when the input to the gate is high i.e., $\theta_a=\pi$, up spin 
conductance disappears for the entire band of energy $E$ 
(Fig.~\ref{notcond}(b)) which reveals that for this typical case electron 
conduction doesn't take place through the magnetic quantum ring. This 
disappearing nature of up spin conductance can be implemented as follows. 
The magnetic moment at the site $\alpha$ is fixed at an angle $\pi$, and 
thus as we tune $\theta_a$ to $\pi$ by applying an external gate magnetic 
filed i.e., the input signal is high, both upper and lower arms of the 
ring are exactly similar in nature. In this situation the ring contributes
nothing for electron conduction at the typical AB flux $\phi=\phi_0/2$. 
Now if $\theta_a$ is dissimilar from $\theta_{\alpha}$, then the two arms 
will not be identical to each other and then the transmission probability 
should not not vanish. Hence, to get zero transmission probability across
the ring when the input signal is
high, we should tune $\theta_{\alpha}$ properly observing $\theta_a$, and 
vice versa. On the other hand, for the other case where the input signal 
is low i.e., $\theta_a=0$, up spin conductance exhibits resonant peaks 
at some particular energies associated with energy eigenvalues of the
magnetic quantum ring. The fact is that, by making input to zero we 
eventually destroy the symmetry among the upper and lower arms of the
ring, and accordingly, for this low input signal ring allows electrons to
conduct through it. Therefore, for low input signal electron is allowed 
to pass through the ring, while in the case of high input electronic 
transmission is completely blocked. This behavior is associated with 
the traditional NOT gate operation.

\begin{figure}[ht]
{\centering \resizebox*{7.75cm}{9cm}{\includegraphics{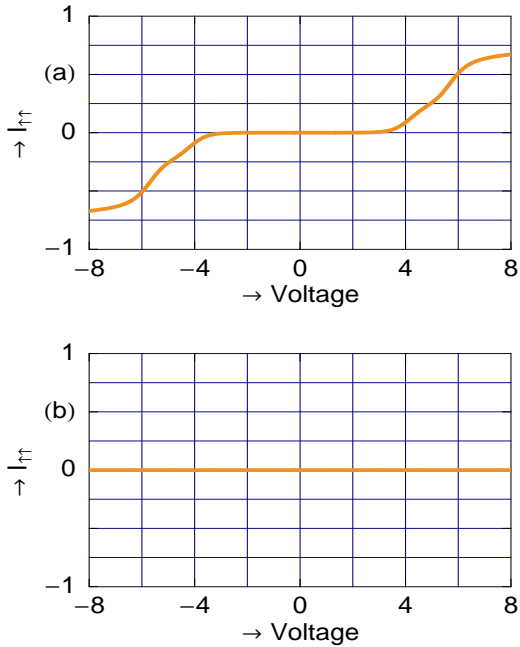}}\par}
\caption{(Color online). NOT gate response. $I_{\uparrow \uparrow}$-$V$ 
curves for a magnetic quantum ring with $N=8$ and $\theta_{\alpha}=\pi$
in the limit of strong-coupling. (a) $\theta_a=0$ and (b) $\theta_a=\pi$. 
Other parameters are as follows: $\phi=\phi_0/2$, $\theta_n=0$ for all 
magnetic sites $n$ except $n=a$ and $\alpha$, and $\varphi_n=0$ for all 
sites $n$.}
\label{notcurr}
\end{figure}

To illustrate the current-voltage characteristics now we concentrate 
on the results given in Fig.~\ref{notcurr}. The up spin currents 
$I_{\uparrow \uparrow}$ are drawn as a function of applied bias voltage 
$V$ for a magnetic quantum ring with $N=8$ in the strong-coupling limit,
where (a) and (b) represent the results for two choices of the input 
signal $\theta_a$. When $\theta_a=\pi$ i.e., high input, up spin current 
vanishes throughout the bias voltage $V$. The result is given in 
Fig.~\ref{notcurr}(b) and for this input signal the vanishing nature 
is clearly followed from the conductance-energy spectrum shown in 
Fig.~\ref{notcond}(b). While, for the case of low input i.e.,
$\theta_a=0$, current gets a finite value (Fig.~\ref{notcurr}(a)) 
following the conductance spectrum (Fig.~\ref{notcond}(a)). Here, the 
current becomes non-zero beyond a threshold voltage $V_{th}$ which is 
tunable depending on the ring size and ring-electrode coupling strength. 
\begin{table}[ht]
\begin{center}
\caption{NOT gate behavior in the limit of strong-coupling. The current
$I$ is computed at the bias voltage $6.26$.}
\label{nottable}
~\\
\begin{tabular}{|c|c|}
\hline \hline
Input ($\theta_a$) & Current ($I$) \\ \hline 
$0$ & $0.562$ \\ \hline
$\pi$ & $0$ \\ \hline \hline
\end{tabular}
\end{center}
\end{table}
From these current-voltage curves it is clear that a high output current 
appears only if the input to the gate is low, while for high input current
doesn't appear. It justifies NOT gate response in the magnetic quantum 
ring.

In a similar way, as we have studied earlier in other logic gates, in 
Table~\ref{nottable} we make a quantitative measurement of the typical 
current amplitude for the magnetic quantum ring. The current amplitude 
is computed at the bias voltage $V=6.26$. It provides that the current 
gets the value $0.562$ when input is low, while it (current) goes to zero 
as we set the input to the high value. Thus the NOT gate operation by 
using a magnetic quantum ring is established.

Up to now we have studied three primary logic gate operations using
one (OR and NOT) and two (AND) magnetic quantum rings. In the forthcoming
sub-sections we will explore the other four combinatorial logic gate
operations using such one or two magnetic quantum rings.

\subsection{NOR gate}

\begin{figure}[ht]
{\centering \resizebox*{7.7cm}{2.9cm}{\includegraphics{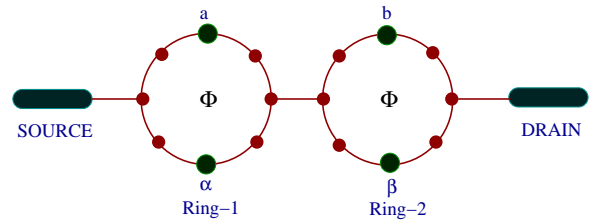}}\par}
\caption{(Color online). Two directly coupled magnetic quantum rings, 
where each ring is penetrated by an AB flux $\phi$, is attached 
symmetrically to two semi-infinite $1$D non-magnetic metallic electrodes.
The magnetic sites $a$, $b$, $\alpha$ and $\beta$ are subject to external
magnetic fields through which the orientations of local magnetic moments 
in these four sites are controlled.}
\label{nor}
\end{figure}

Let us begin with NOR gate operation. Like an AND gate, here we also use
two identical magnetic quantum rings to design a NOR gate. The model 
quantum system is schematically shown in Fig.~\ref{nor}, where two 
magnetic quantum rings, namely, ring-1 and ring-2, are directly coupled 
to each other through a single bond and individual rings are threaded 
by an AB flux $\phi$. The double quantum ring is then attached symmetrically
to two semi-infinite $1$D metallic electrodes, viz, source and drain.
In the upper and lower arms of these two rings we choose four magnetic 
sites referred as $a$, $b$, $\alpha$ and $\beta$ where external magnetic
fields are applied to tune the orientations $\theta_a$, $\theta_b$, 
$\theta_{\alpha}$ and $\theta_{\beta}$ of local magnetic moments in these 
four respective sites. Keeping $\theta_{\alpha}$ and $\theta_{\beta}$ to 
specific values, we regulate $\theta_a$ and $\theta_b$ properly to achieve
NOR gate operation and we call these two ($\theta_a$ and $\theta_b$) as
the two inputs of our NOR gate. Quite nicely we establish that, {\em at 
the typical AB flux $\phi=\phi_0/2$, a high output current ($1$) (in the 
logical sense) appears if both the inputs to the gate are low ($0$), while 
if one or both are high ($1$), a low output current ($0$) results.} This 
phenomenon is the so-called NOR gate response and we will illustrate it by 
describing conductance-energy and current-voltage characteristics.

\begin{figure}[ht]
{\centering \resizebox*{7.75cm}{8cm}{\includegraphics{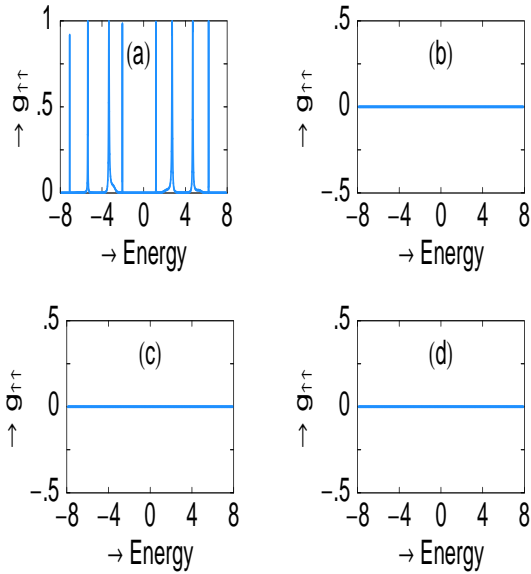}}\par}
\caption{(Color online). NOR gate response. $g_{\uparrow \uparrow}$-$E$ 
curves in the strong-coupling limit for a double quantum ring with 
$\theta_{\alpha}=\theta_{\beta}=\pi$, where each ring contains $8$ 
magnetic sites. 
(a) $\theta_a=\theta_b=0$, (b) $\theta_a=0$ and $\theta_b=\pi$, 
(c) $\theta_a=\pi$ and $\theta_b=0$ and (d) $\theta_a=\theta_b=\pi$. 
Other parameters are as follows: $\phi=\phi_0/2$, $\theta_n=0$ for all 
magnetic sites $n$ except $n=a$, $b$, $\alpha$ and $\beta$, and 
$\varphi_n=0$ for all sites $n$.}
\label{norcond}
\end{figure}

As illustrative examples in Fig.~\ref{norcond} we present the variations
of up spin conductances $g_{\uparrow \uparrow}$ as a function of injecting
electron energy $E$ for a double quantum ring in the limit of strong
ring-to-electrode coupling, where (a), (b), (c) and (d) correspond to 
four different cases of two input signals. The ring sizes are fixed
at $N=8$ and the magnetic moments in $\alpha$ and $\beta$ sites are
rotated by an angle $\pi$ with respect to preferred $+Z$ direction, 
and, throughout this logical operation we fix these two moments in 
such a way. From the conductance-energy characteristics we see that, 
up spin conductance contributes nothing when both the two inputs to 
the gate are set as high i.e., $\theta_a=\theta_b=\pi$ 
(Fig.~\ref{norcond}(d)). The vanishing behavior of up spin conductance 
for this particular choice of input signals can be justified as follows. 
Initially, the magnetic moments in sites $\alpha$ and $\beta$ of lower 
arms of ring-1 and ring-2 are fixed at an angle $\pi$.
Hence, by applying external magnetic fields in the sites $a$ and $b$ as
we tune $\theta_a$ and $\theta_b$ to $\pi$, the upper and lower arms of 
individual rings are exactly similar in nature. In this situation individual
rings contribute nothing to the transmission probability at the typical 
flux $\phi=\phi_0/2$ which provides our desired result. A similar kind 
of vanishing nature of up spin conductance is also observed in the other 
two cases where one input is high and other is low. The results are shown 
in Fig.~\ref{norcond}(b) ($\theta_a=0$ and $\theta_b=\pi$) and in 
Fig.~\ref{norcond}(c) ($\theta_a=\pi$ and $\theta_b=0$). In these two 
choices of input signals, the situation is quite different from the 
particular case where both inputs are high. Here, the symmetry is broken
by applying an external magnetic filed which provides non-zero contribution
to the transmission probability, while it (symmetry) is maintained by 
setting one input to low value which gives zero contribution. Therefore, 
the net contribution to the transmission probability becomes zero as the 
two rings are directly coupled to each other. The non-zero value of
up spin conductance is achieved only when the symmetries among upper 
and lower arms of the two rings are broken individually. It takes place
when both the inputs to the gate are low i.e., $\theta_a=\theta_b=0$
(Fig.~\ref{norcond}(a)). Thus, in short we can say that the electron
\begin{figure}[ht]
{\centering \resizebox*{7.75cm}{8cm}{\includegraphics{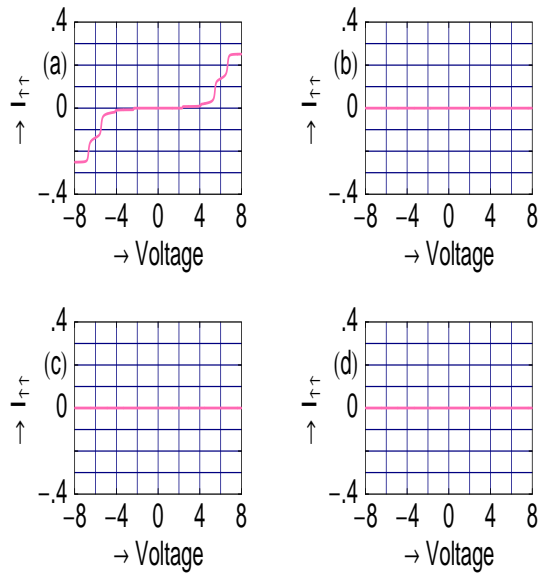}}\par}
\caption{(Color online). NOR gate response. $I_{\uparrow \uparrow}$-$V$ 
curves in the strong-coupling limit for a double quantum ring with 
$\theta_{\alpha}=\theta_{\beta}=\pi$, where each ring contains $8$ 
magnetic sites. 
(a) $\theta_a=\theta_b=0$, (b) $\theta_a=0$ and $\theta_b=\pi$, 
(c) $\theta_a=\pi$ and $\theta_b=0$ and (d) $\theta_a=\theta_b=\pi$. 
Other parameters are as follows: $\phi=\phi_0/2$, $\theta_n=0$ for all 
magnetic sites $n$ except $n=a$, $b$, $\alpha$ and $\beta$, and 
$\varphi_n=0$ for all sites $n$.}
\label{norcurr}
\end{figure}
conduction through the double quantum ring takes place only if both
the inputs to the gate are low, while if anyone or both are high,
electron transmission from the source to drain is forbidden. These
\begin{table}[ht]
\begin{center}
\caption{NOR gate behavior in the limit of strong-coupling. The typical
current amplitude is determined at the bias voltage $V=6.26$.}
\label{nortable}
~\\
\begin{tabular}{|c|c|c|}
\hline \hline
Input-I ($\theta_a$) & Input-II ($\theta_b$) & Current \em{(I)} \\ \hline
$0$ & $0$ & $0.145$ \\ \hline 
$0$ & $\pi$ & $0$ \\ \hline 
$\pi$ & $0$ & $0$ \\ \hline
$\pi$ & $\pi$ & $0$ \\ \hline
\end{tabular}
\end{center}
\end{table}
features agree well with conventional NOR gate operation.

To support the NOR gate operation now we focus our mind on the 
current-voltage characteristics. As representative examples, in
Fig.~\ref{norcurr} we display the variations of up spin currents 
$I_{\uparrow \uparrow}$ with bias voltage $V$ for a double quantum ring 
choosing $\theta_{\alpha}=\theta_{\beta}=\pi$, in the limit of strong 
ring-to-electrode coupling, where (a), (b), (c) and (d) correspond
to four different cases of two input signals $\theta_a$ and $\theta_b$.
From the current-voltage spectra we observe that up spin current 
disappears for the full width of bias voltage $V$ when either both the 
two inputs to the gate are high (Fig.~\ref{norcurr}(d)) or one is high
and other is low (Figs.~\ref{norcurr}(b)-(c)). This vanishing character
of up spin current is followed from the conductance-energy spectra as 
illustrated in Figs.~\ref{norcond}(b)-(d). The non-vanishing feature of
the current is available only when both the two inputs are low 
(Fig.~\ref{norcurr}(a)), obeying the conductance-energy curve plotted 
in Fig.~\ref{norcond}(a). At much low bias voltage, current is almost
zero and it shows a finite value beyond a threshold voltage $V_{th}$
depending on the ring size and ring-to-electrode coupling strength.
These features establish the NOR gate response.

For the sake of our completeness, in Table~\ref{nortable} we do a
quantitative measurement of typical current amplitude, determined at the 
bias voltage $V=6.26$, for the four different choices of two input 
signals in the limit of strong-coupling. Our measurement shows that 
the current gets a finite value ($0.145$) only when both inputs are low
($0$). On the other hand, the current becomes zero for all other cases
i.e., if one or both inputs are high. Therefore, it is manifested that
a double quantum ring can be used as a NOR gate.

\subsection{XOR gate}

As a follow up, now we address XOR gate response which is designed by 
using a single magnetic quantum ring. The ring, penetrated by an AB flux 
\begin{figure}[ht]
{\centering \resizebox*{6.5cm}{3.5cm}{\includegraphics{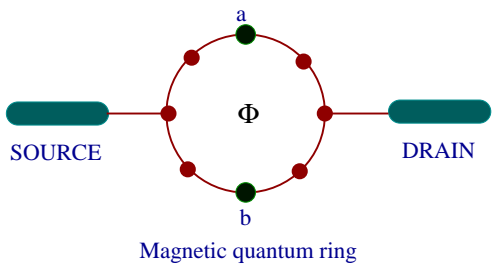}}\par}
\caption{(Color online). A magnetic quantum ring, threaded by an AB flux
$\phi$, is attached symmetrically to two semi-infinite $1$D non-magnetic
metallic electrodes. The magnetic sites $a$ and $b$ in upper and lower 
arms of the ring are subject to external gate magnetic fields through
which the orientations of local magnetic moments in these two sites 
are altered.}
\label{xor}
\end{figure}
$\phi$, is attached symmetrically to two semi-infinite $1$D non-magnetic 
metallic electrodes, namely, source and drain. Our model quantum system 
is schematically shown in Fig.~\ref{xor}. Two magnetic sites, named as 
$a$ and $b$, in upper and lower arms of the ring are subject to external
magnetic fields through which we can regulate the orientations $\theta_a$
and $\theta_b$ of local magnetic moments in these respective sites, and, 
these two ($\theta_a$ and $\theta_b$) are taken as the two inputs of our 
XOR gate. Very nicely we follow that, {\em at the typical AB flux 
$\phi=\phi_0/2$, a high output current ($1$) (in the logical sense) appears 
if one, and only one, of the inputs to the gate is high ($1$), while if 
both inputs are low ($0$) or both are high ($1$), a low output current 
($0$) results.} This is the so-called XOR gate behavior and we will 
emphasize it according to our earlier prescription.

\begin{figure}[ht]
{\centering \resizebox*{7.75cm}{8cm}{\includegraphics{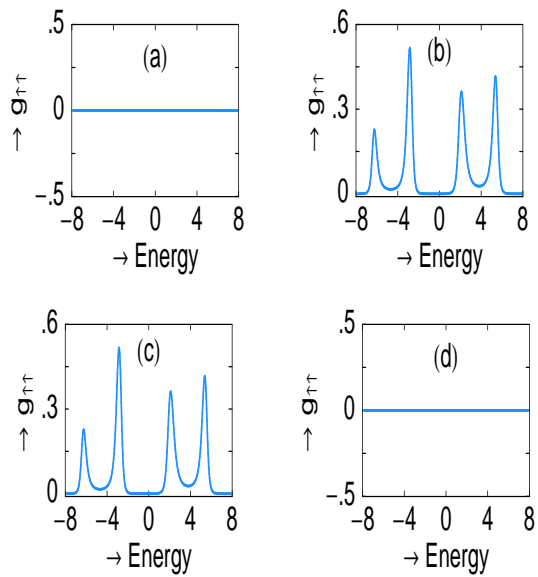}}\par}
\caption{(Color online). XOR gate response. $g_{\uparrow \uparrow}$-$E$ 
curves for a magnetic quantum ring with $N=8$ in the limit of 
strong-coupling. (a) $\theta_a=\theta_b=0$, (b) $\theta_a=0$ and
$\theta_b=\pi$, (c) $\theta_a=\pi$ and $\theta_b=0$ and (d) 
$\theta_a=\theta_b=\pi$. Other parameters are as follows: $\phi=\phi_0/2$, 
$\theta_n=0$ for all magnetic sites $n$ except $n=a$ and $b$, and 
$\varphi_n=0$ for all sites $n$.}
\label{xorcond}
\end{figure}

Let us start with the conductance-energy characteristics given in 
Fig.~\ref{xorcond}. The variations of up spin conductances ($g_{\uparrow 
\uparrow}$) are shown as a function of injecting electron energy $E$ for 
a magnetic quantum ring considering $N=8$ in the strong ring-to-electrode 
coupling limit, where four different figures correspond 
to the results for the different choices of two input signals $\theta_a$ 
and $\theta_b$. It is observed that when both inputs are high i.e.,
$\theta_a=\theta_b=\pi$, up spin conductance gets zero value in the 
complete energy band (see Fig.~\ref{xorcond}(d)). The reason is that
for this typical case both upper and lower arms of the ring are exactly
similar in nature, and therefore, at the AB flux $\phi=\phi_0/2$ 
transmission probability across the rings becomes zero for any energy
$E$ of the source electron. An exactly identical response of up spin
conductance is also visible for the typical case where both inputs are
low i.e., $\theta_a=\theta_b=0$ (Fig.~\ref{xorcond}(a)) and this vanishing
nature can be implemented according to the same prescription as for the 
case high inputs. The non-zero value of up spin conductance is obtained
only when the symmetry among the two arms is broken and it can be done
by setting one input to high and other to low. The results are shown
in Figs.~\ref{xorcond}(b) and (c), where we set $\theta_a=0$, 
$\theta_b=\pi$ and $\theta_a=\pi$, $\theta_b=0$, respectively. Therefore,
from these conductance-energy spectra we can predict that the electron 
can allowed to pass through the magnetic quantum ring provided anyone
\begin{figure}[ht]
{\centering \resizebox*{7.75cm}{8cm}{\includegraphics{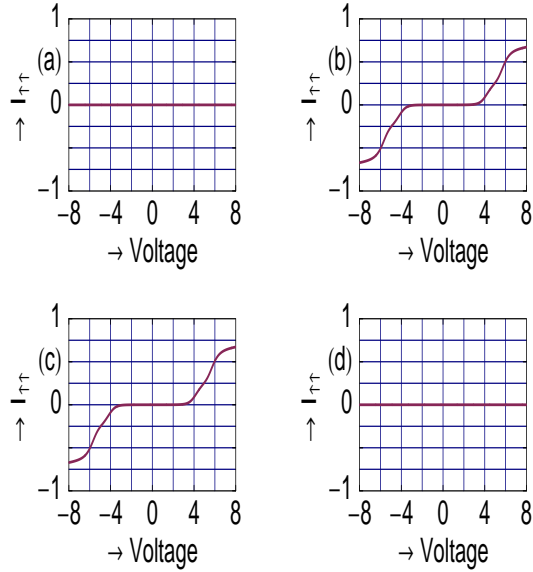}}\par}
\caption{(Color online). XOR gate response. $I_{\uparrow \uparrow}$-$V$ 
curves for a magnetic quantum ring with $N=8$ in the limit of 
strong-coupling. (a) $\theta_a=\theta_b=0$, (b) $\theta_a=0$ and
$\theta_b=\pi$, (c) $\theta_a=\pi$ and $\theta_b=0$ and (d) 
$\theta_a=\theta_b=\pi$. Other parameters are as follows: $\phi=\phi_0/2$, 
$\theta_n=0$ for all magnetic sites $n$ except $n=a$ and $b$, and 
$\varphi_n=0$ for all sites $n$.}
\label{xorcurr}
\end{figure}
input to the gate is high ($\pi$) and other is low ($0$), while for all
other cases i.e., if both inputs are either high or low, transmission of
an electron through the ring becomes forbidden for any energy $E$. 
This phenomenon illustrates the traditional XOR gate response.

With these conductance-energy spectra (Fig.~\ref{xorcond}), now we focus 
our attention on the current-voltage characteristics. As illustrative
\begin{table}[ht]
\begin{center}
\caption{XOR gate behavior in the limit of strong-coupling. The typical
current amplitude is determined at the bias voltage $V=6.26$.}
\label{xortable}
~\\
\begin{tabular}{|c|c|c|}
\hline \hline
Input-I ($\theta_a$) & Input-II ($\theta_b$) & Current \em{(I)} \\ \hline
$0$ & $0$ & $0$ \\ \hline 
$0$ & $\pi$ & $0.562$ \\ \hline 
$\pi$ & $0$ & $0.562$ \\ \hline
$\pi$ & $\pi$ & $0$ \\ \hline
\end{tabular}
\end{center}
\end{table}
purposes in Fig.~\ref{xorcurr} we plot up spin currents $I_{\uparrow 
\uparrow}$ as a function of applied bias voltage $V$ for a typical magnetic 
quantum ring considering $N=8$ in the limit of strong ring-to-electrode 
coupling, where (a)-(d) represent the results for the four different choices
of input signals $\theta_a$ and $\theta_b$. Our results show that up spin 
current contributes nothing as long as both inputs to the gate are set
as low ($0$) or high ($\pi$). For the case of low inputs, result is shown 
in Fig.~\ref{xorcurr}(a), while in Fig.~\ref{xorcurr}(d) the result is 
given when both inputs are high. In these two cases, the vanishing nature 
of the current is justified from our conductance-energy spectra given in 
Figs.~\ref{xorcond}(a) and (d). On the other hand, for other choices of
two input signals current shows non-zero value (Figs.~\ref{xorcurr}(b) 
and (c)), following the conductance spectra (Figs.~\ref{xorcond}(b) and 
(c)). The finite value of up spin current appears when the applied bias
voltage crosses a limiting value, which is the so-called threshold bias
voltage $V_{th}$. Thus to get a current across the ring, we have to take
care about the threshold voltage. These results implement the XOR gate 
response in a magnetic quantum ring.

To make an end of the discussion for XOR gate response in a more compact 
way in Table~\ref{xortable} we make a quantitative measurement of typical 
current amplitude for the four different cases of two input signals. The
current amplitudes are computed at the bias voltage $V=6.26$. It is 
observed that current becomes zero when both inputs are either low or high. 
While, it (current) reaches the value $0.562$ when we set one input as high
and other as low. These studies suggest that a magnetic quantum ring can 
be used as a XOR gate.

\subsection{XNOR gate}

As a consequence now we will explore XNOR gate response and we design this
logic gate by means of a single magnetic quantum ring. The ring, threaded 
by an AB flux $\phi$, is attached symmetrically to two semi-infinite $1$D 
non-magnetic metallic electrodes, namely, source and drain.
\begin{figure}[ht]
{\centering \resizebox*{6.5cm}{3.3cm}{\includegraphics{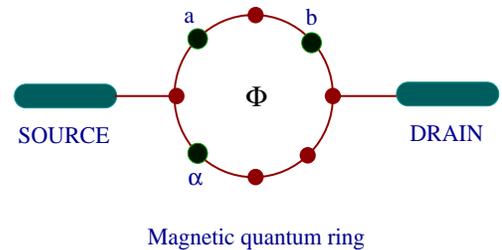}}\par}
\caption{(Color online). A magnetic quantum ring, penetrated by an AB
flux $\phi$, is attached symmetrically to two semi-infinite $1$D 
non-magnetic metallic electrodes, viz, source and drain. The magnetic 
sites $a$, $b$ and $\alpha$ are subject to external gate magnetic fields
through which the orientations of local magnetic moments in these three
sites are changed.}
\label{xnor}
\end{figure}
The model quantum system is schematically shown in Fig.~\ref{xnor}. 
Two magnetic sites specified as $a$ and $b$ situated in the upper arm and 
one magnetic site named as $\alpha$ placed in the lower arm of the ring 
are subject to external gate magnetic fields through which the orientations 
$\theta_a$, $\theta_b$ and $\theta_{\alpha}$ of local magnetic moments in 
the respective sites $a$, $b$ and $\alpha$ can be altered. These $\theta_a$ 
and $\theta_b$ are considered as the inputs of our two-input XNOR logic gate.  
We show that, {\em at the typical magnetic flux $\phi=\phi_0/2$ a high output 
current ($1$) (in the logical sense) appears if both the two inputs to the 
gate are the same, while if one but not both inputs are high ($1$), a low 
output current ($0$) results.} This logical operation is the so-called
XNOR gate behavior and we will focus it by studying conductance-energy
spectrum and current-voltage characteristics for a typical magnetic quantum
ring.

\begin{figure}[ht]
{\centering \resizebox*{7.75cm}{8cm}{\includegraphics{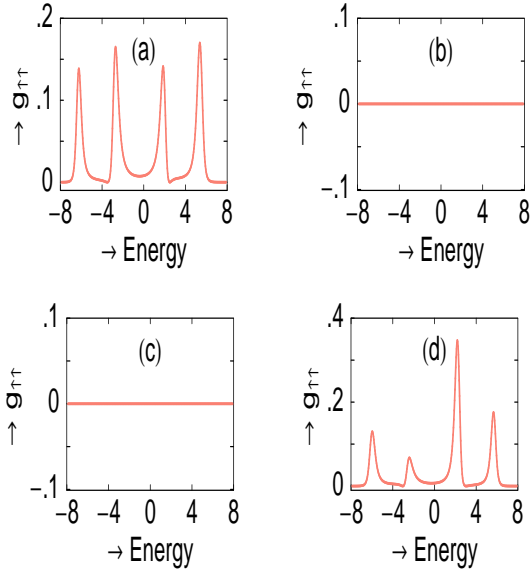}}\par}
\caption{(Color online). XNOR gate response. $g_{\uparrow \uparrow}$-$E$ 
curves for a magnetic quantum ring with $N=8$ and $\theta_{\alpha}=\pi$
in the limit of strong-coupling. (a) $\theta_a=\theta_b=0$, (b) 
$\theta_a=0$ and $\theta_b=\pi$, (c) $\theta_a=\pi$ and $\theta_b=0$ 
and (d) $\theta_a=\theta_b=\pi$. Other parameters are as follows: 
$\phi=\phi_0/2$, $\theta_n=0$ for all magnetic sites $n$ except $n=a$, $b$,
and $\alpha$, and $\varphi_n=0$ for all sites $n$.}
\label{xnorcond}
\end{figure}

As representative examples, in Fig.~\ref{xnorcond} we show the variations 
of up spin conductances $g_{\uparrow \uparrow}$ as a function of injecting 
electron energy $E$ for a typical magnetic quantum ring with $N=8$ and 
$\theta_{\alpha}=\pi$ in the limit of strong ring-to-electrode coupling, 
where (a)-(d) correspond to the
four different cases of two input signals $\theta_a$ and $\theta_b$.
From the results it is noticed that for the cases where one input is
high ($\pi$) and other is low ($0$) i.e., $\theta_a=0$, $\theta_b=\pi$ 
and $\theta_a=\pi$, $\theta_b=0$, up spin conductance disappears for the 
entire energy band (Figs.~\ref{xnorcond}(b) and (c)), and therefore, for 
these two choices of input signals electronic transmission through the 
magnetic quantum ring is completely forbidden. This feature can be 
implemented as follows. The magnetic moment in site $\alpha$ is 
rotated by an angle $\pi$ and throughout this logic gate operation it
is aligned in such a way. Thus, as we set anyone input to high ($\pi$) 
by applying an external gate magnetic field and other to low ($0$), upper 
and lower arms of the ring become exactly identical in nature to each 
other which provide zero transmission probability at the typical AB flux 
$\phi=\phi_0/2$. For any other orientation of the moment placed at the
site $\alpha$ i.e., if $\theta_{\alpha}\ne \pi$ the vanishing transmission 
probability will not appear for the cases where one input is set at a 
high value and other is fixed to a low value. Hence, to achieve zero
transmission probability across the bridge, we have to fix $\theta_{\alpha}$ 
properly considering the input signals and vice versa. For the other two 
cases of input signals i.e., when both the inputs are either low 
($\theta_a=\theta_b=0$) or high ($\theta_a=\theta_b=\pi$), non-vanishing
transmission probability of up spin electrons is observed. In these two 
particular cases, up spin conductance shows resonant peaks 
(Figs.~\ref{xnorcond}(a) and (d)) for some typical energies associated 
with the energy eigenvalues of the magnetic quantum ring. This 
non-vanishing nature of up spin conductance is quite obvious, since 
\begin{figure}[ht]
{\centering \resizebox*{7.75cm}{8cm}{\includegraphics{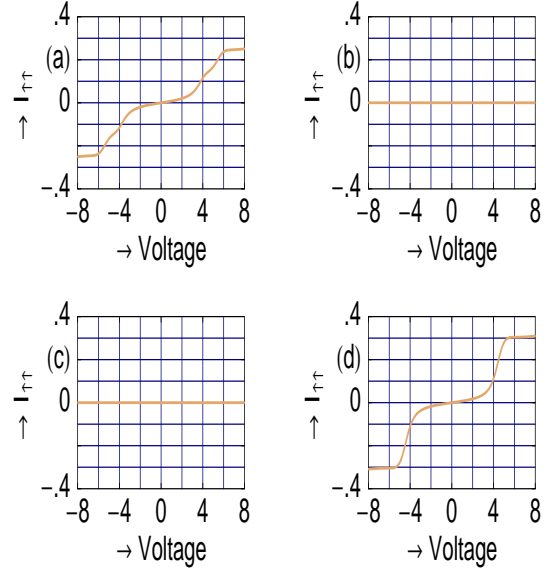}}\par}
\caption{(Color online). XNOR gate response. $I_{\uparrow \uparrow}$-$V$ 
curves for a magnetic quantum ring with $N=8$ and $\theta_{\alpha}=\pi$
in the limit of strong-coupling. (a) $\theta_a=\theta_b=0$, (b) 
$\theta_a=0$ and $\theta_b=\pi$, (c) $\theta_a=\pi$ and $\theta_b=0$ 
and (d) $\theta_a=\theta_b=\pi$. Other parameters are as follows: 
$\phi=\phi_0/2$, $\theta_n=0$ for all magnetic sites $n$ except $n=a$, $b$,
and $\alpha$, and $\varphi_n=0$ for all sites $n$.}
\label{xnorcurr}
\end{figure}
in these two cases of input signals symmetry between two arms of the 
magnetic quantum ring is broken. It is also observed that the heights 
of the resonant peaks are very small compared to unity which is due 
to the effect of quantum interferences among the two arms of the ring.
From these conductance-energy spectra of four different choices of input
signals it can be manifested that the electronic conduction through the 
magnetic quantum ring is possible only when both inputs to the gate are 
either low ($0$) or high ($\pi$). While, for all other cases electron 
conduction through the ring is no longer possible. This phenomenon 
reveals the conventional XNOR gate response.

As a continuation now we follow the current-voltage characteristics to
reveal XNOR gate response. In Fig.~\ref{xnorcurr} we plot the variations
of up spin currents $I_{\uparrow \uparrow}$ as a function of applied bias
voltage $V$ for a typical magnetic quantum ring with $N=8$ and 
$\theta_{\alpha}=\pi$ in the strong-coupling limit, where (a)-(d) 
correspond to the results for the four different choices of input 
signals $\theta_a$ and $\theta_b$. Following the conductance-energy
spectra shown in Figs.~\ref{xnorcond}(b) and (c), we see that up spin
current drops to zero for any bias voltage $V$ when anyone input is
fixed at a high value and other is kept at low. The results are
presented in Figs.~\ref{xnorcurr}(b) and (c). 
\begin{table}[ht]
\begin{center}
\caption{XNOR gate behavior in the limit of strong-coupling. The typical
current amplitude is determined at the bias voltage $V=6.26$.}
\label{xnortable}
~\\
\begin{tabular}{|c|c|c|}
\hline \hline
Input-I ($\theta_a$) & Input-II ($\theta_b$) & Current \em{(I)} \\ \hline
$0$ & $0$ & $0.243$ \\ \hline 
$0$ & $\pi$ & $0$ \\ \hline 
$\pi$ & $0$ & $0$ \\ \hline
$\pi$ & $\pi$ & $0.304$ \\ \hline
\end{tabular}
\end{center}
\end{table}
The finite contribution of up current is available only when the symmetry 
between the two arms of the ring is broken either by making two inputs 
zero i.e., $\theta_a=\theta_b=0$ (Fig.~\ref{xnorcurr}(a)), or by applying 
magnetic fields to the input gates i.e., $\theta_a=\theta_b=\pi$ 
(Fig.~\ref{xnorcurr}(d)). These current-voltage characteristics justify 
the XNOR gate response in a magnetic quantum ring. 

To be more precise, in Table~\ref{xnortable} we make a quantitative 
measurement of typical current amplitude for the different choices of
two input signals in the strong ring-to-electrode coupling. The typical 
current amplitude is computed at the bias voltage $V=6.26$. It is noticed 
that current gets the value $0.243$ when both inputs are low ($0$), while 
it becomes $0.304$ when both inputs to the gate are high ($\pi$). On the
other hand for all other cases, current is always zero. These results 
emphasize that a magnetic quantum ring can be used to design a XNOR gate.

\subsection{NAND gate}

\begin{figure}[ht]
{\centering \resizebox*{6.5cm}{3.3cm}{\includegraphics{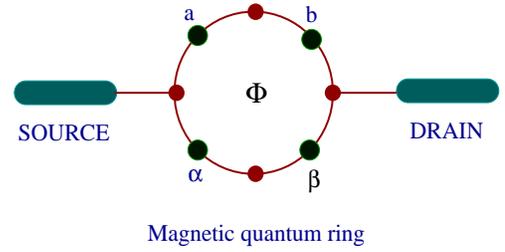}}\par}
\caption{(Color online). A magnetic quantum ring, threaded by an AB flux
$\phi$, is attached symmetrically to two semi-infinite $1$D non-magnetic
metallic electrodes, namely, source and drain. The magnetic sites $a$, 
$b$, $\alpha$ and $\beta$ are subject to external gate magnetic fields
through which the directions of local magnetic moments in these four 
sites are controlled.}
\label{nand}
\end{figure}

At the end, we demonstrate NAND gate response and we design this logic
gate with the help of a single magnetic quantum ring. The ring, penetrated
by an AB flux $\phi$, is attached symmetrically to two semi-infinite $1$D 
non-magnetic metallic electrodes, viz, source and drain. The schematic
view of the magnetic quantum ring that can be used to design a NAND gate
is shown in Fig.~\ref{nand}. In the lower arm of the ring, two magnetic 
sites labeled as $\alpha$ and $\beta$ are subject to external magnetic 
fields through which the orientations $\theta_{\alpha}$ and $\theta_{\beta}$ 
of magnetic moments in these respective sites are altered. Throughout
the NAND gate operation we set $\theta_{\alpha}=\theta_{\beta}=\pi$. In 
addition, we choose another two sites specified as $\alpha$ and $\beta$
in upper arm of the ring where external magnetic fields are applied 
through which the orientations $\theta_a$ and $\theta_b$ of magnetic 
moments in these two sites are controlled. These two ($\theta_a$ and 
$\theta_b$) are variable and they are considered as two input of the 
NAND gate. Quite interestingly we notice that, {\em at the typical AB 
flux $\phi=\phi_0/2$ a high output current ($1$) (in the logical sense) 
appears if one or both inputs to the gate are low ($0$), while if both 
inputs to the gate are high ($1$), a low output current ($0$) results.} 
\begin{figure}[ht]
{\centering \resizebox*{7.75cm}{8cm}{\includegraphics{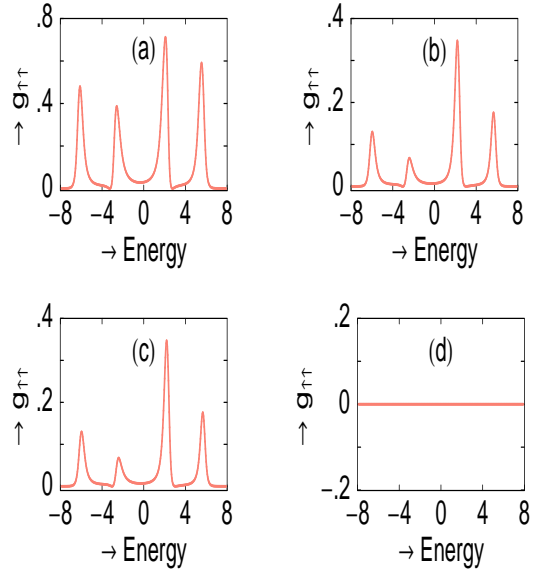}}\par}
\caption{(Color online). NAND gate response. $g_{\uparrow \uparrow}$-$E$ 
curves for a magnetic quantum ring with $N=8$ and $\theta_{\alpha}
=\theta_{\beta}=\pi$ in the limit of strong-coupling. 
(a) $\theta_a=\theta_b=0$, (b) $\theta_a=0$ and $\theta_b=\pi$, (c) 
$\theta_a=\pi$ and $\theta_b=0$ and (d) $\theta_a=\theta_b=\pi$. Other 
parameters are as follows: $\phi=\phi_0/2$, $\theta_n=0$ for all magnetic 
sites $n$ except $n=a$, $b$, $\alpha$ and $\beta$, and $\varphi_n=0$ for 
all sites $n$.}
\label{nandcond}
\end{figure}
This characteristic is the so-called NAND gate response and we will 
justify it by describing the conductance-energy and current-voltage
spectra.

Let us begin with the results given in Fig.~\ref{nandcond}. Here we 
show the variations of up spin conductances $g_{\uparrow \uparrow}$ as 
a function of injecting electron energy $E$ for a typical magnetic 
quantum ring with $N=8$ in the limit of strong ring-to-electrode
coupling considering $\theta_{\alpha}=\theta_{\beta}=\pi$, where (a)-(d)
correspond to four different choices of two input signals $\theta_a$ and 
$\theta_b$. Our results predict that for the typical case when both the 
two inputs to the gate are high i.e., $\theta_a=\theta_b=\pi$, up spin 
conductance disappears for the complete energy band which reveals no 
electronic transmission through the magnetic quantum ring 
(Fig.~\ref{nandcond}(d)). The reason is that, for this particular choice
of two inputs, both the upper and lower arms of the ring become exactly 
identical in nature, and therefore, at the typical flux $\phi=\phi_0/2$
the ring contributes nothing to the electronic transmission probability. 
On the other hand, for all other possible cases of two input signals 
i.e., either when both the two inputs are low ($0$) or anyone is low
\begin{figure}[ht]
{\centering \resizebox*{7.75cm}{8cm}{\includegraphics{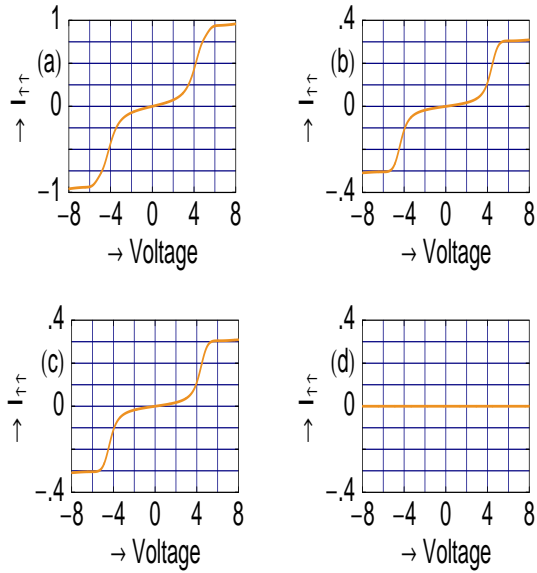}}\par}
\caption{(Color online). NAND gate response. $I_{\uparrow \uparrow}$-$V$ 
curves for a magnetic quantum ring with $N=8$ and $\theta_{\alpha}
=\theta_{\beta}=\pi$ in the limit of strong-coupling. 
(a) $\theta_a=\theta_b=0$, (b) 
$\theta_a=0$ and $\theta_b=\pi$, (c) $\theta_a=\pi$ and $\theta_b=0$ 
and (d) $\theta_a=\theta_b=\pi$. Other parameters are as follows: 
$\phi=\phi_0/2$, $\theta_n=0$ for all magnetic sites $n$ except $n=a$, $b$,
$\alpha$ and $\beta$, and $\varphi_n=0$ for all sites $n$.}
\label{nandcurr}
\end{figure}
and other is high, up spin conductance shows resonant peaks for some 
particular energies associated with the energy eigenvalues of the magnetic 
quantum ring. The results are presented in Figs.~\ref{nandcond}(a)-(c). 
In these three cases, symmetry among the two arms of the ring is no longer 
exists, and therefore, non-zero transmission probability appears. A careful 
observation predicts that the heights of resonant peaks for the cases where 
only one input is high and other is low (Figs.~\ref{nandcond}(b)-(c)) get 
reduced significantly compared to the case where both inputs are low 
(Fig.~\ref{nandcond}(a)). This is solely due to the effect of quantum 
interference among the two arms of the magnetic quantum ring. Thus,
from these conductance-energy spectra we can predict that the electron 
conduction through the ring takes place if one or both the inputs are low
($0$), while if both are high ($\pi$) no electron conduction takes place
across the bridge system. These results justify the traditional NAND gate 
operation.

In the same fashion, now we focus on the current-voltage characteristics. 
As our illustrative purposes, in Fig.~\ref{nandcurr} we present the 
variations of up spin currents $I_{\uparrow \uparrow}$ in terms of
applied bias voltage $V$ for a typical magnetic quantum ring with 
$N=8$ in the limit of strong-coupling,
\begin{table}[ht]
\begin{center}
\caption{NAND gate behavior in the limit of strong-coupling. The typical
current amplitude is determined at the bias voltage $V=6.26$.}
\label{nandtable}
~\\
\begin{tabular}{|c|c|c|}
\hline \hline
Input-I ($\theta_a$) & Input-II ($\theta_b$) & Current \em{(I)} \\ \hline
$0$ & $0$ & $0.939$ \\ \hline 
$0$ & $\pi$ & $0.304$ \\ \hline 
$\pi$ & $0$ & $0.304$ \\ \hline
$\pi$ & $\pi$ & $0$ \\ \hline
\end{tabular}
\end{center}
\end{table}
considering $\theta_{\alpha}=\theta_{\beta}=\pi$, where (a)-(d) represent
the results for the different cases of two input signals. When both inputs
are high i.e, $\theta_a=\theta_b=\pi$ up spin current becomes zero for the 
entire range of applied bias voltage (Fig.~\ref{nandcurr}(d)), following
the conductance-energy spectrum given in Fig.~\ref{nandcond}(d) since the
current is determined by integrating the transmission function. For all 
the other three choices of two inputs, finite contribution in up spin 
current is available. The results are shown in Figs.~\ref{nandcurr}(a)-(c). 
From these three current-voltage spectra, it is observed that for a fixed 
bias voltage current amplitude in the typical case where both inputs are
low (Fig.~\ref{nandcurr}(a)) is much higher than the cases where one
input is high and other is low (Figs.~\ref{nandcurr}(b)-(c)). This is
clearly understood from the variations of conductance-energy spectra
studied in the above paragraph. Thus, our present current-voltage 
characteristics justify the NAND gate operation in the magnetic quantum
ring very nicely.

Finally, in Table~\ref{nandtable} we present a quantitative estimate of
the typical current amplitude for the four different cases of two input 
signals. The typical current amplitudes are measured at the bias voltage 
$V=6.26$. It provides that the current vanishes when both inputs are high 
($\pi$). On the other hand, the current gets the value $0.939$ as long as 
both inputs are low ($0$) and $0.304$ when anyone of two inputs is low 
and other is high. Our results support that a magnetic quantum ring can 
be utilized as a NAND gate. 

\section{Concluding remarks}

In the present work, we have implemented classical logic gates like OR, 
AND, NOT, NOR, XOR, XNOR and NAND at nano-scale level using magnetic 
quantum rings. A single ring is used to design OR, NOT, XOR, XNOR and
NAND gates, while the rest two gates are fabricated by using two such
rings and in all the cases each ring is penetrated by an AB flux $\phi$
which plays the crucial role for the whole logic gate operations. We 
have used a simple tight-binding framework to describe the model, where
a magnetic quantum ring is attached to two semi-infinite one-dimensional
non-magnetic metallic electrodes. Based on a single particle Green's
formalism all the calculations have been done numerically which 
demonstrate two-terminal conductance and current through the system. 
Our theoretical analysis may be useful in fabricating mesoscopic or
nano-scale logic gates.

Throughout our work, we have studied seven possible logic gates. Out of 
which five logic gates are designed by using a single magnetic quantum 
ring, while the rest two are fabricated with the help of two magnetic
quantum rings. In the case of single rings, we have chosen the rings with 
total number of atomic sites $N=8$. On the other hand, for the coupled 
ring systems, we have considered two identical rings, where each ring 
contains $8$ atomic sites. In our model calculations, these typical 
numbers ($8$ or $2\times8=16$) are chosen only for the sake of simplicity. 
Though the results presented here change numerically with the ring size 
($N$), but all the basic features remain exactly invariant. To be more 
specific, it is important to note that, in real situation the experimentally
achievable rings have typical diameters within the range $0.4$-$0.6$
$\mu$m. In such a small ring, unrealistically very high magnetic fields
are required to produce a quantum flux. To overcome this situation,
Hod {\em et al.} have studied extensively and proposed how to construct
nanometer scale devices, based on Aharonov-Bohm interferometry, those
can be operated in moderate magnetic fields~\cite{baer4,baer5,baer6,baer7}.

In the present paper we have done all the calculations by ignoring
the effects of temperature, electron-electron correlation, disorder,
etc. Due to these factors, any scattering process that appears in the
mesoscopic ring would have influence on electronic phases, and, in
consequences can disturb the quantum interference effects. Here we
have assumed that, in our samples all these effects are too small, and
accordingly, we have neglected all these factors in this particular 
study.

The importance of this article is mainly concerned with (i) the simplicity 
of the geometry and (ii) the smallness of the size. 

\vskip 0.2in
\noindent
{\bf\small ACKNOWLEDGMENTS}
\vskip 0.2in
\noindent
I acknowledge with deep sense of gratitude the illuminating comments
and suggestions I have received from Prof. S. N. Karmakar and Moumita 
Dey during the calculations.

\end{document}